\documentclass[fleqn,usenatbib]{mnras}

\usepackage{newtxtext,newtxmath}

\usepackage[T1]{fontenc}
\usepackage{ae,aecompl}
\usepackage{multirow}
\usepackage{hyperref}

\DeclareRobustCommand{\VAN}[3]{#2}
\let\VANthebibliography\thebibliography
\def\thebibliography{\DeclareRobustCommand{\VAN}[3]{##3}\VANthebibliography}

\usepackage{graphicx}	
\usepackage{amsmath}	

\usepackage{amssymb}	
\usepackage{comment}
\usepackage{bm}
\usepackage{ulem}

\def\ee{\end{equation}}
\def\be{\begin{equation}}

\def\gpcyr{\rm \,Gpc^{-3}\,yr^{-1}}
\def\Rmrgbh{R_{\bullet\bullet}}
\def\Rmrge{R_{\bullet\bullet,\rm E}}
\def\Rmrgd{R_{\bullet\bullet,\rm D}}

\def\mc{{{\mathcal{M}}}_{\rm c}}

\title[Multiband GW observations]{Multiband gravitational wave observations of steller binary black holes at the low to middle and high frequencies}

\author[Zhao et al.]{Yuetong Zhao$^{1,2}$\thanks{ytzhao@nao.cas.cn}, Youjun Lu$^{1,2}$\thanks{luyj@nao.cas.cn}, Changshuo Yan$^{1,2}$\thanks{yancs@nao.cas.cn}, 
Zhiwei Chen$^{1,2}$, and Wei-Tou Ni$^{1,3}$\\
$^{1}$\,CAS Key Laboratory for Computational Astrophysics, National Astronomical Observatories, Chinese Academy of Sciences,  20A Datun Road, Beijing 100101, China\\
$^{2}$\,School of Astronomy and Space Sciences, University of Chinese Academy of Sciences, 19A Yuquan Road, Beijing 100049, China \\
$^{3}$\,Innovation Academy of Precision Measurement Science and Technology (APM), Chinese Academy of Sciences, Wuhan 430071, China
}

\date{Accepted XXX. Received YYY; in original form ZZZ}

\pubyear{2023}

\begin{document}
\label{firstpage}
\pagerange{\pageref{firstpage}--\pageref{lastpage}}
\maketitle

\begin{abstract}
The ground-based gravitational wave (GW) observatories discover a population of merging stellar binary black holes (BBHs), which are promising targets for multiband observations by the low-, middle-, and high-frequency GW detectors. In this paper, we investigate the multiband GW detections of BBHs and demonstrate the advantages of such observations in improving the localization and parameter estimates of the sources. We generate mock samples of BBHs by considering different formation models as well as the merger rate density constrained by the current observations (GWTC-3). We specifically consider the astrodynamical middle-frequency interferometer GW observatory (AMIGO) in the middle-frequency band and estimate that it may detect $21$-$91$ BBHs with signal-to-noise ratio $\varrho\geq8$ in a $4$-yr observation period. 
The multiband observations by the low-frequency detectors [Laser Interferometer Space Antenna (LISA) and Taiji] and AMIGO may detect $5$-$33$ BBHs with $\varrho_{\rm LT}\geq5$ and $\varrho_{\rm AMI}\geq5$, which can evolve to the high-frequency band within $4$ yr and can be detected by the Cosmic Explorer (CE) and Einstein Telescope (ET). The joint observations of LISA-Taiji-AMIGO-ET-CE may localize the majority of the detectable BBHs in sky areas of $7\times10^{-7}$ to $2\times10^{-3}$\,deg$^2$, which is improved by a factor of $\sim120$, $\sim2.4\times10^{5}$, $\sim1.8\times10^{4}$, or $\sim1.2\times10^{4}$, comparing with those by only adopting CE-ET, AMIGO, LISA-Taiji, or LISA-Taiji-AMIGO. These joint observations can also lead to an improvement of the measurement precision of the chirp mass (symmetric mass ratio) by a factor of $\sim5.5\times10^{4}$ ($33$), $\sim16$ ($8$), $\sim120$ ($90$), or $\sim5$ ($5$), comparing with those by CE-ET, AMIGO, LISA-Taiji, or LISA-Taiji-AMIGO.
\end{abstract}




\section{Introduction}

LVK (LIGO/Virgo/KAGRA) collaborations report the detection of high-frequency ($\sim10-300$\,Hz) gravitational waves (GWs) emitted from $90$ compact binary mergers, including more than $80$ stellar mass binary black hole (hereafter BBH) mergers, two binary neutron star (BNS) mergers and four neutron star black hole (NSBH) mergers \citep{GWTC3catalog, GW150914, GW190521}. Such compact binaries may also emit GWs at lower frequencies ($\sim 10^{-3}-10 \rm{Hz}$) during their inspiral stage before the merging stage, and thus may be detected by GW detectors at lower frequencies, such as the Laser Interferometer Space Antenna (LISA), \url{https://lisa.nasa.gov}, Taiji \citep{SSPMAHuang2017}, TianQin \citep{Tianqin2016}, the astrodynamical middle-frequency interferometer GW observatory (AMIGO, \citealt{Ni2018}), the DECi-hertz Interferometer Gravitational wave Observatory (DECIGO, \citealt{decigo}), B-DECIGO \citep{bdecigo}, and the Big Bang Observer (BBO, \citealt{bbo2006}). 
Therefore, multiband GW observations may be realized with future GW detectors constructed in different frequency bands \citep[][]{Sesana2016}, such as the high-frequency ground-based GW detectors [e.g., the Cosmic Explorer (CE, \citealt{3rdGWD}), and the Einstein Telescope (ET, \citealt{ETstudy})], the milli-Hertz and deci-Hertz space GW detectors (e.g., LISA, Taiji, and AMIGO).

Multiband GW observations of compact binaries may provide complementary information of these binaries, with which one may not only obtain refined source parameter estimations but also significant information about the formation and evolution of these sources that the observations at a single band cannot provide \citep[e.g.,][]{Sesana2016, 2017LISA, Gerosa2019, LiuShao2020, Izumi2021, ChenLu2021, Klein2022}. For example, the BBHs formed via the dynamical interactions of stars in dense stellar systems or AGN discs may have significant eccentricities when they emit GWs in the low-frequency band but are well circularized in the high-frequency band \citep{Benacquista2002, Rodriguez2016, 2016ApJ...830L..18B, Samsing2018, D'Orazio2018, Mapelli2022, Antonini2012, Stephan2016, Bartos2017, AGNecc2018ApJ, Samsing2022Nat}, while those formed via the evolution of massive binary stars in the galactic field at the low-frequency band are expected to be on orbits close to circular \citep{Cris2002, Cris2016, Mandel2016, Son2020}. Therefore, the combination of the low-frequency and high-frequency GW observations can provide a probe to distinguish different stellar BBH formation mechanisms. Multiband observations may also provide measurements of the post-Newtonian phasing coefficients to accuracy about one to two orders of magnitude better than the bounds achievable from the GW signals detected by single detectors (either CE/ET in the high-frequency band or LISA in the low-frequency band) \citep{2020PhRvL.125t1101G}. The detection of BNSs by decihertz GW observatories may provide early-warning parameters for both the high-frequency GW detection and electromagnetic counterpart searching \citep[e.g.,][]{2022ApJ...934...84L, 2022MNRAS.515..739K}.

Heavy BBHs are probably the most important sources for multiband (milli-Hertz to hecto-Hertz) GW observations. \citet{Sesana2016} first pointed out the importance of multiband observations for BBHs and estimated that around hundreds of BBHs resolved by LISA would merge in the LIGO band in $10$ years. In principle, these sources are also the main targets for the decihertz GW detectors. The joint observations of the BBHs by the low-frequency, middle-frequency, and high-frequency GW detectors should further help to improve the parameter estimations as well as constraining their origins, etc. Multiband GW studies in the literature mainly focus on the joint observations of either BBHs by the millihertz and hectohertz GW detectors \citep{Sesana2016, 2017LISA, Gerosa2019, LiuShao2020, Izumi2021, ChenLu2021, Klein2022} or BNS by the decihertz and hectohertz GW detectors \citep{2022ApJ...934...84L, 2022MNRAS.515..739K}. Recently, \citet{Isoyama2018} investigated the multiband observations by LISA, B-DECIGO, and hectohertz GW detectors for GW150914 and GW170817 like systems. In this paper, we investigate the multiband GW joint observations of BBHs by millihertz, decihertz, and hectohertz GW detectors (CE/ET). We estimate the detectable number of BBHs (jointly) by these GW detectors and illustrate the importance of the joint multiband GW observations in improving the parameter estimations of these sources.

This paper is organized as follows. In Section~\ref{sec:sample}, we describe the mock samples of BBHs originated from different formation models. We introduce the method to estimate the signal-to-noise (S/N) ratio for the mock sources, and then estimate the expected numbers and parameters distributions of the ``detectable''\footnote{Here ``detectable'' with quote marks means those mock BBHs that can be detected with  S/N above the specifically adopted threshold for each GW detector or network under consideration. In the following text, we omit the quote marks for simplicity.} BBHs (jointly) by different GW detectors in Section~\ref{sec:snr}. In Section~\ref{sec:PE}, we adopt the Fisher information matrix (FIM) method to estimate the measurement precision for the physical and geometrical parameters of those detectable BBHs detected either by single GW detectors or by multiband GW networks. Discussions and conclusions are presented in Section~\ref{sec:dis_con}.

Throughout this paper, we adopt the standard $\Lambda$CDM cosmology model with $H_0 = 67.9 {\rm km s^{-1} Mpc^{-1}}$, $\Omega_{\rm m} = 0.306$, $\Omega_{\rm k} = 0$, and $\Omega_{\Lambda} = 0.694$ \citep{Planck2016}.

\section{Mock Sample of BBHs}
\label{sec:sample}

In order to investigate the prospects of multiband observations of GWs from BBHs evolving from inspiral stage to the final merger stage, we adopt four different models regarding their formation channels as well as the observational constraint from LVK collaboration \citep{GWTC3population}. We generate mock samples of BBHs for each model by the following number distribution with respect to the GW frequency in the observer's rest frame $f_{\rm o}$ \citep[see][]{ZhaoLu2021} 
\begin{eqnarray}
\label{eq:num-circ}
\frac{dN}{df_{\rm o}} \simeq \frac{dN}{dt_{\rm o}}\cdot \frac{dt_{\rm o}}{df_{\rm o}} = \iiint \Rmrgbh(z)P(m_1,q) \times \nonumber \\
\frac{1}{(1+z)} \frac{dV}{dz} \frac{dt_{\rm o}}{df_{\rm o}} dm_1 dq dz,
\end{eqnarray}
where $\Rmrgbh(z)$ is the merger rate density at redshift $z$, $P(m_1,q)$ is the source parameter distribution ($m_1$, $m_2$, and $q = m_2/m_1$ representing the primary mass, the secondary mass, and the mass ratio of BBHs), $dV/dz = (4\pi c D_{\rm c}^2)/(H_0\sqrt{\Omega_m(1+z)^3+\Omega_\Lambda})$ is the comoving volume element, $c$ is the speed of light, $D_{\rm c}$ is the comoving distance of sources, and $dt_{\rm o}/df_{\rm o} = (5/96)\pi^{-8/3}(G\mc/c^3)^{-5/3}f_{\rm o}^{-11/3}$ is the differential residential time of BBHs in the observer's frame with $\mc = (1+z)m_1q^{3/5}(1+q)^{-1/5}$ representing the redshifted chirp mass and $G$ the gravitational constant. We only consider GW sources with $f_{\rm o}$ ranging from $0.003$ to $1$\,Hz since the coalescence time-scale is longer than $100$ yr at lower frequency even for heavy BBHs. Regarding different formation channels of BBHs, the redshift evolution of merger rate density and underlying chirp mass distribution are different. Our settings for these models are detailed below. Here, we generate mock BBHs for millihertz and decihertz GW detectors which is different from the method for generating mock samples in hectohertz detectors as in the work by \citet{Chenzw2022}.

\begin{itemize}
\item {\bf EMBS model:} In this model, we assume all the BBHs are originated from the evolution of massive binary stars in galactic field (hereafter denoted as the EMBS channel). Their merger rate density $\Rmrge$ is calculated by the convolution of their birth rate with the time delay ($t_{\rm d}$) distribution \citep{Dvorkin2016,CaoLu2018}. The BBHs' birth rate density is derived from the cosmic star formation rate (SFR) $\dot{\psi}(Z;z)$ at redshift $z$ with metallicity $Z$ \citep{Madau2014,Cris2016} and the initial mass function (IMF) $\phi(m_\star)$ \citep{Salpeter1955} by
\be
R_{\rm birth}(m_1,z) = \iint dm_\star dZ \dot{\psi}(Z;
z)\phi(m_\star) \delta (m_\star - g^{-1}(m_1,Z)),
\label{eq:Rbir}
\ee
where $m_1 = g(m_\star, Z)$ is the remnant mass and progenitor stellar mass relationship given by \citet{Spera2015}. As for the time delay of the BBH merger from its birth, we adopt a distribution $P(t_{\rm d})\propto t_{\rm d}^{-1}$ the same as the work of \citet{Cris2016}. More detailed descriptions can be found in section $2.2$ in \citet{ZhaoLu2021}.

\item {\bf Dynamical model:} Here, we assume all BBHs are formed through dynamical interactions in a dense stellar environments such as globular clusters (GCs; hereafter denoted as the Dynamical model). The merger rate density evolution can be obtained as in \citet{Mapelli2022} by using dynamical simulations of the formation of BBHs in GCs and simple descriptions of the formation of GCs. The formation rate of GCs is assumed to follow a Gaussian distribution
\be
\psi_{\rm GC}(z) = \mathcal{B}_{\rm GC}e^{-(z-z_{\rm GC})^2/(2\sigma_{\rm GC}^2)},
\label{eq:Rgc}
\ee
where $z_{\rm GC} = 3.2$ is the peak redshift of formation for GCs, $\sigma_{\rm GC} = 1.5$, and $\mathcal{B}_{\rm GC} = 2\times10^{-4}{M_\odot \rm Mpc^{-3}yr^{-1}}$ is the fiducial normalization factor \citep{Mapelli2022}, which is reminiscent of those in \citet{ElBadry2019} and \citet{Rodriguez2018}. Then the BBHs merger rate density is given by
\begin{eqnarray}
\Rmrgd(z) = \int_{z_{\rm max}}^{z}\frac{\psi_{\rm GC}(z^{\prime})}{(1+z^{\prime})H_0\sqrt{\Omega_{\rm m}(1+z)^3+\Omega_\Lambda}}\times  \nonumber \\
\left[\int_{Z_{\rm min}(z^{\prime})}^{Z_{\rm max}(z^{\prime})}\eta(Z)\mathcal{F}(z^{\prime}, z, \rm Z)dZ\right]dz^{\prime}.
\end{eqnarray}
Here $\mathcal{F}(z^{\prime},z, Z)$ is the merger rate of BBHs that form at redshift $z^{\prime}$ from progenitors with metallicity $Z$ and merge at redshift $z$ in globular clusters, $\eta(Z)$ is the merger efficiency at metallicity $Z$, $Z_{\rm min}(z^{\prime})$, and $Z_{\rm max}(z^{\prime})$ are the minimum and maximum metallicity of progenitor stars formed at redshift $z^{\prime}$. We adopt the $\rm A03$ GC channel model in \citet[see their table for details]{Mapelli2022}. The resulting BBHs from the dynamical model tend to be relatively more massive than those from the EMBS model because of the hierarchical mergers in the dynamical model.

\item{\bf Hybrid model:} Considering both the EMBS formation channel and the dynamical formation channel, where $75\%$ of the BBHs are assumed to be from the EMBS channel while the rest from the dynamical channel. This setting was also illustrated in \citet{ZhaoLu2021}.

\item{\bf GWTC-3 model:} This model is an observationally constrained model according to the latest results from LVK collaborations \citep{GWTC3population}. The redshift dependent merger rate density is proportional to $(1+z)^{\kappa}$ with $\kappa = 2.7^{+1.8}_{-1.9}$ at low redshift range(i.e. $z \leq 1$). We adopt the fiducial power law plus peak (Power Law + Peak (PP)) model for primary mass and mass ratio distribution in \citet{GWTC3population}. We consider this model because almost all BBHs detectable jointly by the multiband GW detectors are at redshfit $z<1$.
\end{itemize}

The local merger rate density for each of the first three models is normalized to the latest constraint provided by LVK after the first three observational runs (i.e., $\Rmrgbh(0) = 19.1^{+8.4}_{-8.5} \gpcyr$) \citep{GWTC3population}. We generate mock BBHs via the Monte Carlo method according to the redshift evolution of the merger rate density, the primary mass distribution, and mass ratio distributions of BBHs in each model. Considering the uncertainty in the local merger rate density constrained by LVK, we generate $100$ realizations of mock samples for each model with three different values of $\Rmrgbh(0)$, i.e., $19.1$, $10.6$, and $27.5 \gpcyr$, indicating the median value as well as $90\%$ credible intervals of the constrained local merger rate density.

\section{Detectability of BBHs for Different GW detectors}
\label{sec:snr}

According to the matched filtering method \citep{Finn1992}, the S/N ratio of each GW source can be estimated as
\be
\label{eq:snr}
\varrho^2 = (h|h) = \int_{f_{\rm i}}^{f_{\rm f}} \frac{4\tilde{h}^*(f)\tilde{h}(f)}{S_{\rm n}(f)}df,  
\ee
where $\tilde{h}(f)$ is the Fourier transform of GW signal $h(t)$ from the source, $S_{\rm n}(f)$ is the non-sky-averaged spectral strain sensitivity of a single detector, $f_{\rm i}$ is the initial frequency assigned via the Monte Carlo generation of mock BBHs at the beginning of the observation, and $f_{\rm f} = {\rm min}(f_{\rm end}, f_{\rm ISCO}, f_{\rm detector})$. Here, $f_{\rm end}$ represents the frequency of the GW emitted by the BBH at the end of the observation,  $f_{\rm ISCO} = 2.2M_{\odot}/(m_1(1+q)(1+z))\, \rm{kHz}$ is the GW frequency of the BBH when it is at the innermost circular stable orbit, $f_{\rm detector}$ denotes the upper frequency limit of a certain GW detector. The S/N of a source detected by multiple GW detectors or a GW network can be thus estimated as
\be
\label{eq:snrmulti}
\varrho^2 = \sum_{j=1}^{n}\int_{f_{\rm i}}^{f_{\rm f}}\frac{4\tilde{h}^*_j(f)\tilde{h}_j(f)}{S_{{\rm n},j}(f)}df,
\ee
where $j$ represents independent detectors and $n$ refers to the total number of Michelson interferometers in the detector network. For the low-frequency GW detectors, we adopt the LISA sensitivity curve from \citet{Robson2019} and the Taiji ones from \citet{Wanggang2020}. We specifically take account of the middle frequency GW detector AMIGO by using the three different levels of sensitivity proposed in \citet{Ni2022}. 

BBHs in the inspiral stage will emit GWs within the frequency range of LISA/Taiji/AMIGO, and their GW strains can be approximated as \citep{Maggiore2008}:
\be
\label{eq:hf}
\tilde{h}(f) = \left(\frac{5}{24}\right)^{1/2}\frac{1}{\pi^{2/3}}\frac{c}{D_{\rm L}}\left(\frac{G \mc}{c^3}\right)^{5/6}f^{-7/6}e^{i \Psi(f)}Q,
\ee
where $D_{\rm L}$ is the luminosity distance, and $Q$ and $\Psi(f)$ are two quantities defined below. The  response of GW signals to the detectors is reflected in the quantity $Q$ as
\be
\label{eq:Q}
Q = \sqrt{F_+^2\left(\frac{1+\cos^2\iota}{2}\right)^2+F_{\times}^2\cos^2\iota},
\ee
where $F_+$ and $F_{\times}$ are the detector's pattern functions as 
\be
\label{eq:F_+}
F_+ = \frac{1}{2} \left(1+\cos^2 \theta \right) \cos 2\phi \cos 2\psi - \cos \theta \sin 2\phi \sin 2\psi,
\ee
and
\be
\label{eq:F_cross}
F_{\times} = \frac{1}{2}\left(1+\cos^2 \theta \right) \cos 2\phi \sin 2\psi + \cos \theta \sin 2\phi \cos 2\psi.
\ee
In equations~\eqref{eq:Q}-\eqref{eq:F_cross}, $\theta$, $\phi$, and $\psi$ are the polar angle, azimuthal angle, and polarization angle of the stellar BBH in the detector's frame and $\iota$ is the inclination angle between the BBH angular momentum and the vector pointing from the detector to the BBH. $\Psi(f)$ is the phase evolution of GW strain with consideration of the polarization modulation ($\phi_{\rm p}$) and a Doppler phase correction ($\phi_{\rm D}$), i.e.,
\begin{eqnarray}
\label{eq:psi_f}
\Psi(f) = 2\pi f t_{\rm c} -\phi_{\rm c}-\frac{\pi}{4}-\phi_{\rm p}-\phi_{\rm D}+\frac{3}{4}(8\pi\mc f)^{-5/3}  \nonumber \\
\times \left[1+\frac{20}{9}\left(\frac{743}{336}+\frac{11\eta}{4}\right)x-16\pi x^{3/2} \right].
\end{eqnarray}
Here $t_{\rm c}$ and $\phi_{\rm c}$ are the coalescence time and orbital phase at coalescence, $x(f) = [\pi m_1(1+q)(1+z) f]^{2/3}$ is the PN parameter, and $\eta = q/(1+q)^2$ is the symmetric mass ratio. The polarization modulation is defined as 
\be
\label{eq:phip}
\phi_{\rm p} = \arctan \frac{-2 \cos \iota F_{\times}(t(f))}{(1+\cos^2 \iota)F_{+}(t(f))}.
\ee
The Doppler phase correction caused by the orbital motion of the space-based 
detectors around the Sun is \citep{Cutler1998}
\be
\label{eq:phiD}
\phi_{\rm D}(t(f)) = 2\pi f(t) R \sin\theta_{\rm S} \cos \left(\Phi(t(f))-\phi_{\rm S} \right), 
\ee
where $R = 1$\,au is the semimajor axis of the detector, $\Phi(t) = \Phi_0+\frac{2\pi t(f)}{T}$ is the azimuthal angle of the detector orbiting the Sun with $T$ representing the detector's orbital period, and $t(f)$ can be expressed as 
\be
\label{eq:tf}
t(f) = t_{\rm c}-5(8\pi f)^{-8/3}\times\left[1+\frac{4}{3} \left(\frac{743}{336}+\frac{11\eta}{4}\right)x-\frac{32\pi}{5}x^{3/2} \right].
\ee
When calculating the non-sky-averaged S/N we transform these parameters ($\theta$, $\phi$, $\psi$) from the detector's frame to the ecliptic coordinate frame ($\theta_{\rm S}$, $\phi_{\rm S}$, $\theta_{\rm L}$, $\phi_{\rm L}$) (see e.g. equations 3.16-3.22 in \citealt{Cutler1998}). To generate the mock samples, we assume that $\phi_{\rm S}$,$\phi_{\rm L}$ are uniformly distributed in [$0$, $2\pi$), $\cos\theta_{\rm S}$ and $\cos\theta_{\rm L}$ are uniformly distributed in [$-1$, $1$]. The S/N estimates also depend on the initial phase $\Phi_0$ and the detector arm's initial orientation angle $\alpha_0$. 

For the high frequency ($1$ - $1000 \rm{Hz}$) ground-based GW detectors, we adopt the IMRPhenomD waveform model in {\bf PyCBC} \citep{2019PASP..131b4503B} to calculate S/N according to Equation~\eqref{eq:snr}. For simplicity, we only consider non-spinning, circular BBHs in this paper.

We generate $100$ realizations of BBHs at redshift $z \leq 0.8$ with the initial frequency ranging from $0.003$\,Hz to $1$\,Hz for each model with $3$ different local merger rate densities (i.e., $10.6$, $19.1$, and $27.5\gpcyr$), respectively. We calculate the expected S/N of each stellar BBH observed by LISA, Taiji, and AMIGO according to Equation~\eqref{eq:snr}, and that is observed by different combinations of these detectors according to Equation~\eqref{eq:snrmulti}. Throughout this work, we assume that the continuous observational periods for LISA, Taiji, and AMIGO are all $4$ yrs and a stellar BBH is detectable if the expected S/N in each frequency band is larger than a threshold $\varrho_{\rm th} = 8$. When considering multiband observations, we assume a stellar BBH is a genuinely detectable multiband source only if the expected S/Ns for each of the considered frequency band by specific detector/network (e.g., LISA-Taiji (LT), AMIGO, ET-CE) are all larger than $5$. Here, we simply assume that those GW detectors start their observation mission at the same time regardless of real observation plans. We rank the number of detectable BBHs from low to high among $100$ realizations of each model. The $50$th, $16$th, and $84$th ranks are adopted to represent the median value and the $68\%$ confidence interval of the predicted number of detectable BBHs. Note that here we do not specially discuss the detectable BBHs for the high-frequency GW detectors, such as ET and CE, as they can all be detected by ET and CE with high S/N.

\begin{figure}
\includegraphics[width=\columnwidth]{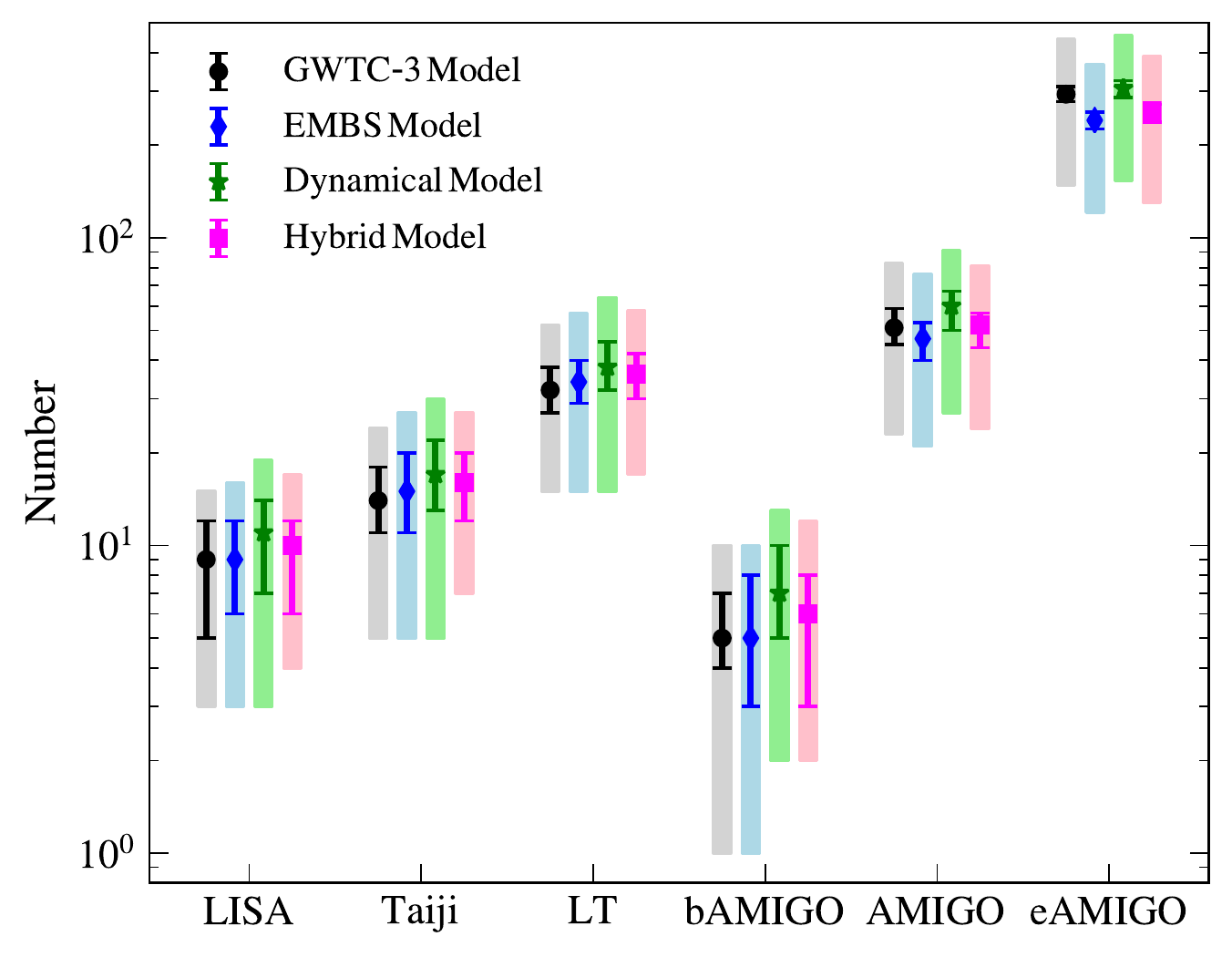}
\caption{Estimated numbers of detectable BBHs with S/N $\geq 8$ by LISA, Taiji, LT, bAMIGO, AMIGO, and eAMIGO within an observation period of $4$ yr for different models, respectively. The filled circle/diamond/star/square symbols with errorbars represent the median value and the $68\%$ confidence interval of the numbers resulting from $100$ realizations for the GWTC-3, EMBS, dynamical, and hybrid models, respectively, where the local merger rate density is set to $\Rmrgbh = 19.1\gpcyr$. 
The grey shaded region associated with each symbol indicates the range induced by the local merger rate density uncertainty, with the lower/upper bound representing the $16$th/$84$th (detectable number ranking from small to large) among $100$ realizations with $\Rmrgbh = 10.6/27.5\gpcyr$.
}
\label{fig:f1}
\end{figure}

Figure~\ref{fig:f1} shows the detectability of BBHs for LISA, Taiji, {\bf LT,} and AMIGO with three different sensitivity levels (labelled as bAMIGO, AMIGO, and eAMIGO) under four different models. As seen from Figure~\ref{fig:f1}, the expected numbers of detectable BBHs by each detector are not so sensitive to the choice of the model, though the dynamical model gives the highest values for all the detectors because of the relatively larger chirp masses of BBHs produced by this model comparing with those by other models \citep{ZhaoLu2021}. The uncertainty in the estimates of these numbers is mainly inherited from the uncertainty in the constraint on the local merger rate density. We find that LISA alone is able to detect $\sim 3$-$19$ BBHs in $4$ yr, Taiji alone can detect $\sim 5$-$30$ BBHs, slightly more than those by LISA because of its better sensitivity at the high frequency (see  Fig.~\ref{fig:f12}), and LT can detect $\sim 15$-$64$ BBHs, a few times more than those by either LISA or Taiji \citep[see also][]{ChenLu2021}. Detecting the inspiralling BBHs is one of the main scientific goals of the middle-frequency GW detectors like bAMIGO/AMIGO/eAMIGO \citep{Ni2022}.
For bAMIGO, it can only detect $\sim 1$-$13$ BBHs in $4$ yr. AMIGO and eAMIGO can detect $\sim 21$-$91$ and $\sim 121$-$454$ BBHs in $4$ yr, substantially larger than those by LISA, Taiji, and bAMIGO. Table~\ref{tab:number} lists the numbers of the detectable BBHs by each of these detectors resulting from different models with the assumption of $R_{\bullet\bullet,0}=19.1\gpcyr$ and the lower/upper bound for the observational constrained merger rate density being $10.6/27.5\gpcyr$.

\begin{figure*}
\includegraphics[scale = 0.6]{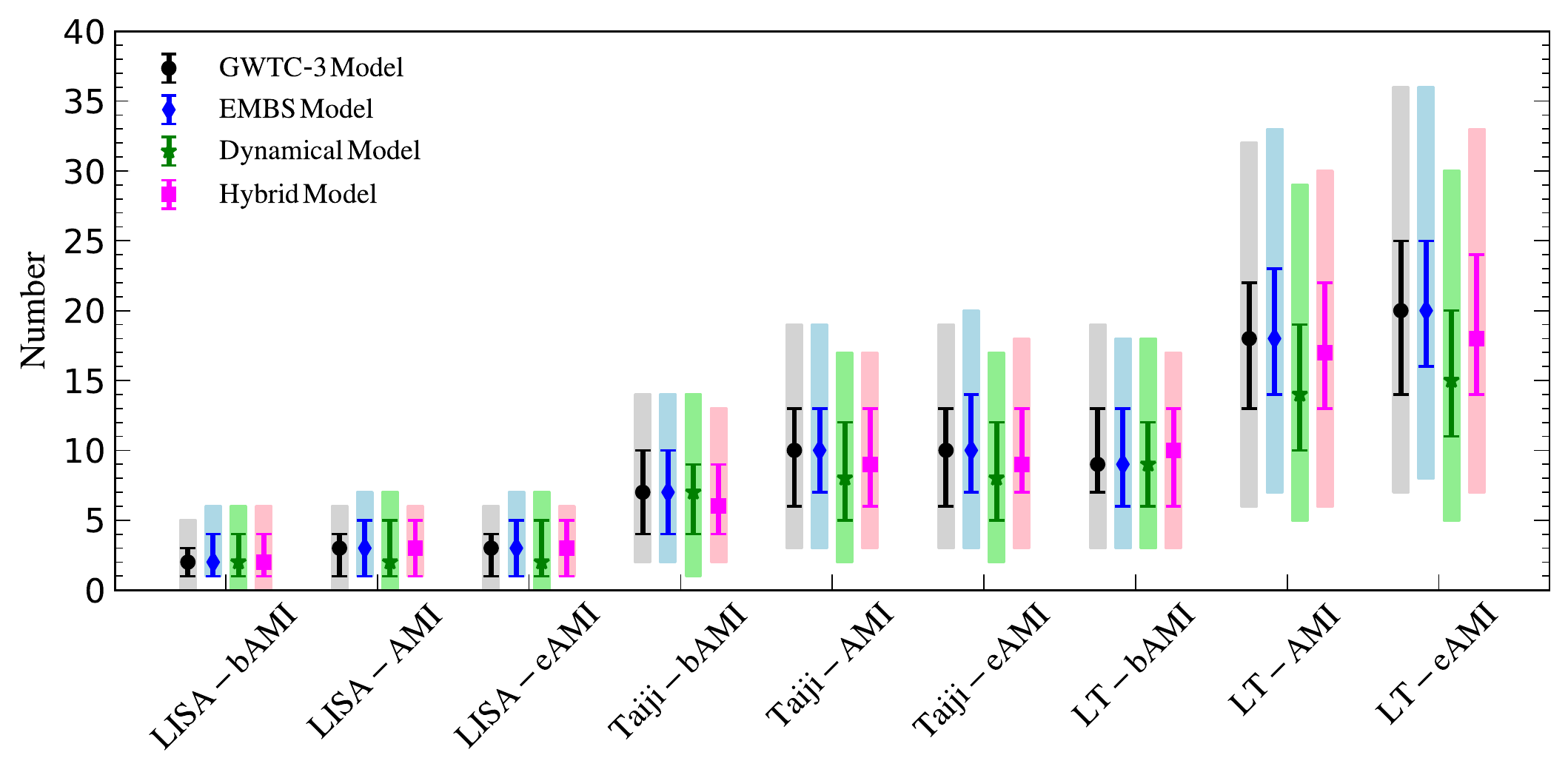}
\caption{Legends are the same as that for Fig.~\ref{fig:f1}, except that for different combinations of multiband GW detectors. Here, `LT' represents the LISA-Taiji network, b-AMIGO, AMIGO, and e-AMIGO are labelled as `bAMI', `AMI', and `eAMI', respectively.
}
\label{fig:f2}
\end{figure*}

\begin{table*}
\centering
\caption{
Expected number of detectable BBH events from different models by $4$ yr observations of LISA, Taiji, LT, bAMIGO, AMIGO, and eAMIGO, and those of multiband BBH events by joint observations of detectors in the low-frequency and middle-frequency bands. 
}
\label{tab:number}
\resizebox{1\linewidth}{!}{
\begin{tabular}{c|c|c|c|c|c|c|c|c|c|c|c|c} %
\hline		
\hline
\multirow{2}{*}{GW detector}  & \multicolumn{3}{c}{GWTC-3 model}  & \multicolumn{3}{c}{EMBS model}  & \multicolumn{3}{c}{Dynamical model}  & \multicolumn{3}{c}{Hybrid model}  \\  
& ${\rm low}\Rmrgbh$ & ${\rm mid}\Rmrgbh$ & ${\rm high}\Rmrgbh$ & ${\rm low}\Rmrgbh$ & ${\rm mid}\Rmrgbh$ & ${\rm high}\Rmrgbh$ & ${\rm low}\Rmrgbh$ & ${\rm mid}\Rmrgbh$ & ${\rm high}\Rmrgbh$ & ${\rm low}\Rmrgbh$ & ${\rm mid}\Rmrgbh$ & ${\rm high}\Rmrgbh$ \\ \hline
 {LISA} & $3$ & $9^{+3}_{-4}$ & $15$ & $3$ & $9^{+3}_{-3}$ & $16$ & $3$ & $11^{+3}_{-4}$ & $19$ & $4$ & $10^{+2}_{-4}$ & $17$ \\ \hline
 {Taiji} & $5$ & $14^{+4}_{-3}$ & $24$ & $5$ & $15^{+5}_{-4}$ & $27$ & $5$ & $17^{+5}_{-4}$ & $30$ & $7$ & $16^{+4}_{-4}$ & $27$ \\ \hline
 {bAMIGO} & $1$ &$5^{+2}_{-1}$ & $10$ & $1$ & $5^{+3}_{-2}$ & $10$ & $2$ & $7^{+3}_{-2}$ & $13$ & $2$ & $6^{+2}_{-3}$ & $12$ \\ \hline
 {AMIGO} & $23$ & $51^{+8}_{-6}$ & $83$ & $21$ & $47^{+6}_{-7}$ & $76$ & $27$ & $60^{+7}_{-10}$ & $91$ & $24$ & $52^{+5}_{-7}$ & $81$ \\ \hline
 {eAMIGO} & $148$ & $293^{+17}_{-16}$ & $442$ & $121$ & $241^{+15}_{-15}$ & $365$ & $153$ & $306^{+18}_{-20}$ & $454$ & $130$ & $259^{+13}_{-21}$ & $388$ \\ \hline
 {LT} & $15$ & $32^{+6}_{-5}$ & $52$ & $15$ & $34^{+6}_{-5}$ & $57$ & $15$ & $38^{+8}_{-6}$ & $64$ & $17$ & $36^{+6}_{-6}$ & $58$ \\ \hline
 {LISA-bAMIGO} & $0$ & $2^{+1}_{-1}$ & $5$ & $1$ & $2^{+2}_{-1}$ & $4$ & $0$ & $2^{+2}_{-1}$ & $6$ & $0$ & $2^{+2}_{-1}$ & $6$ \\ \hline
 {LISA-AMIGO} & $0$ & $3^{+1}_{-2}$ & $6$ & $1$ & $3^{+2}_{-2}$ & $7$ & $0$ & $2^{+3}_{-1}$ & $7$ & $1$ & $3^{+2}_{-2}$ & $6$ \\ \hline
 {LISA-eAMIGO} & $0$ & $3^{+1}_{-2}$ & $6$ & $1$ & $3^{+2}_{-2}$ & $7$ & $0$ & $2^{+3}_{-1}$ & $7$ & $1$ & $3^{+2}_{-2}$ & $6$\\ \hline
 {Taiji-bAMIGO} & $2$ & $7^{+3}_{-3}$ & $14$ & $2$ & $7^{+3}_{-3}$ & $14$ & $1$ & $7^{+2}_{-3}$ & $14$ & $2$ & $6^{+3}_{-2}$ & $13$ \\ \hline
 {Taiji-AMIGO} & $3$ & $10^{+3}_{-4}$ & $19$ & $3$ & $10^{+3}_{-3}$ & $19$ & $2$ & $8^{+4}_{-3}$ & $17$ & $3$ & $9^{+4}_{-3}$ & $17$ \\ \hline
 {Taiji-eAMIGO} & $3$ & $10^{+3}_{-4}$ & $19$ & $3$ & $10^{+4}_{-3}$ & $20$ & $2$ & $8^{+4}_{-3}$ & $17$ & $3$ & $9^{+4}_{-2}$ & $18$\\ \hline
 {LT-bAMIGO} & $3$ & $9^{+4}_{-2}$ & $19$ & $3$ & $9^{+4}_{-3}$ & $18$ & $3$ & $9^{+3}_{-3}$ & $18$ & $3$ & $10^{+3}_{-4}$ & $17$ \\ \hline
 {LT-AMIGO} & $6$ & $18^{+4}_{-5}$ & $32$ & $7$ & $18^{+5}_{-4}$ & $33$ & $5$ & $14^{+5}_{-4}$ & $33$ & $6$ & $17^{+5}_{-4}$ & $30$ \\ \hline
 {LT-eAMIGO} & $7$ & $20^{+5}_{-6}$ & $36$ & $8$ & $20^{+5}_{-4}$ & $36$ & $5$ & $15^{+5}_{-4}$ & $30$ & $7$ & $18^{+6}_{-4}$ & $33$ \\ \hline
\end{tabular}
}
\begin{flushleft}
\footnotesize{Note: first column denotes the GW detector name or different combinations of them, here `LT' represents the LISA-Taiji network. Second column shows the $16$th detectable number ranking from small to large of BBHs for GWTC-$3$ model with $\Rmrgbh = 10.6\gpcyr$ labelled as $\rm{low}\Rmrgbh$. Third column shows the median value and $68$ percent confidence interval of the numbers among $100$ realizations of BBHs for GWTC-$3$ model with $\Rmrgbh = 19.1\gpcyr$ labelled as $\rm{mid}\Rmrgbh$. Fourth column shows the $84$th detectable number for GWTC-$3$ model with $\Rmrgbh = 27.5\gpcyr$ labelled as $\rm{high}\Rmrgbh$. The fifth to seventh columns, the eighth to tenth columns and the last three columns show the corresponding results for the EMBS, dynamical, and hybrid models, respectively.}
\end{flushleft}
\end{table*}

Figure~\ref{fig:f2} shows the detectability of multiband BBHs via the joint observations of low-frequency GW detectors (LISA, Taiji, or LT) and middle frequency GW detector (bAMIGO, AMIGO, or eAMIGO).
Considering the uncertainty in the constraint on the local BBH merger rate density, LISA-bAMIGO/LISA-AMIGO/LISA-eAMIGO is expected to detect $\sim 0$-$6$/$0$-$7$/$0$-$7$ BBHs, Taiji-bAMIGO/Taiji-AMIGO/Taiji-eAMIGO is expected to detect $\sim 1$-$14$/$2$-$19$/$2$-$20$ BBHs, and LT-bAMIGO/LT-AMIGO/LT-eAMIGO is expected to detect $\sim 3$-$19$/$5$-$33$/$5$-$36$ BBHs. The uncertainties of these number estimations are mainly due to the uncertainty in the constraint on the local merger rate density. For simplicity, we mainly consider AMIGO among the three levels of sensitivity designs in our following analysis, if not otherwise stated.

\begin{figure*}
\centering
\includegraphics[width=0.8\textwidth]{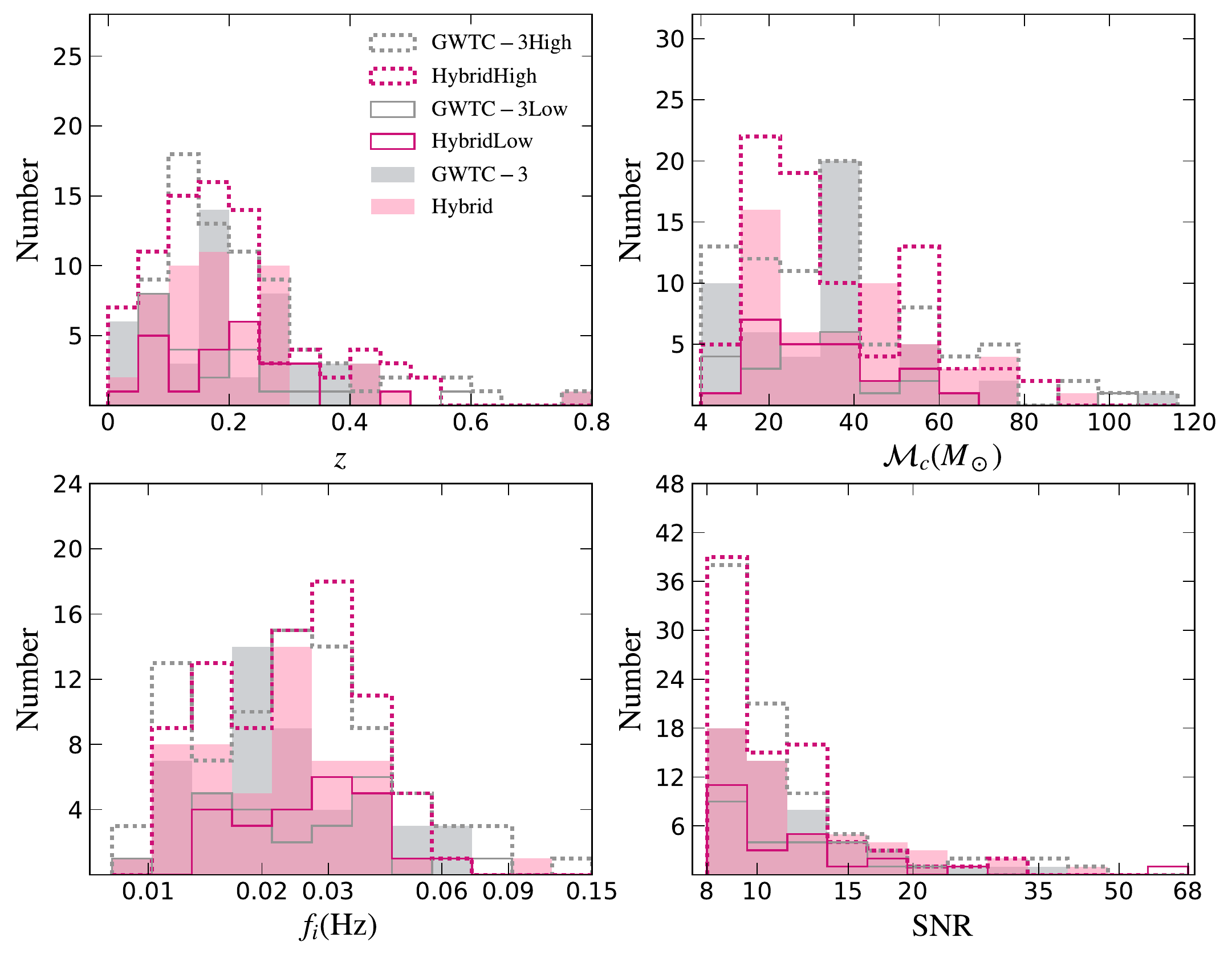}
\caption{The redshift (top left panel), chirp mass (top right panel), initial frequency (bottom left panel) and S/N (bottom right panel) distributions of the detectable mock BBHs by AMIGO for the GWTC-$3$ and hybrid models. The coloured shaded histograms show the results of the realization with the detectable number ranking of $50$th (from small to large) among $100$ realizations conducted for the GWTC-3 (grey) and hybrid (pink) models, respectively, in which the local merger rate density is normalized to $R_{\bullet\bullet,0} = 19.1\gpcyr$. The grey/pink dotted histograms represent the realization with the detectable number ranking as the $84$th among those of the $100$ realizations for the GWTC-$3$/hybrid model with $R_{\bullet\bullet,0} = 27.5\gpcyr$ (GWTC-$3$ High/Hybrid High). Similarly, grey/pink solid histograms represent the one with $16$th rank among the $100$ realizations in the GWTC-$3$/hybrid model with $R_{\bullet\bullet,0} = 10.6\gpcyr$ (GWTC-$3$ Low/Hybrid Low). }
\label{fig:f3}
\end{figure*}

\begin{figure*}
\centering
\includegraphics[width=0.8\textwidth]{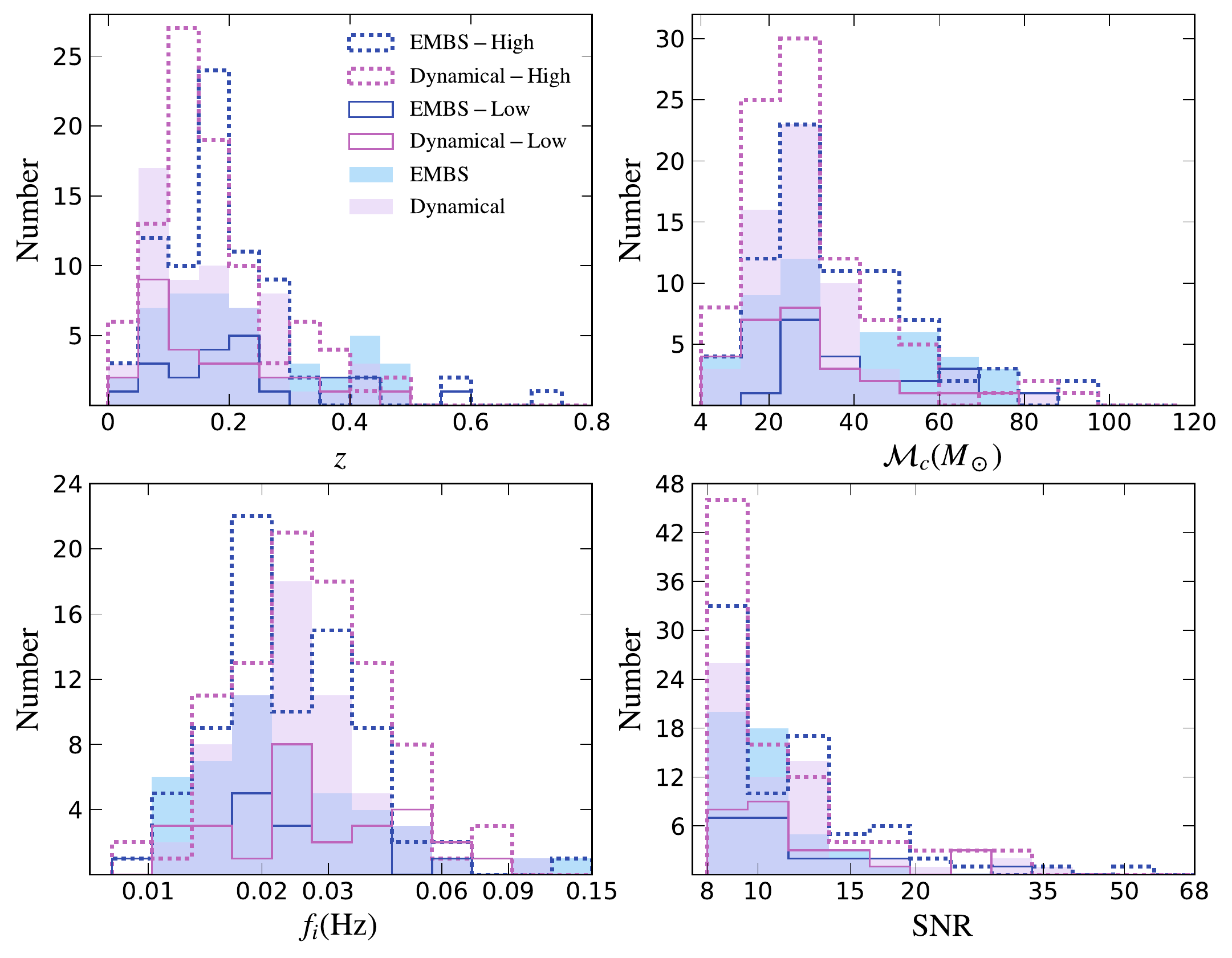}
\caption{The redshift (top left panel), chirp mass (top right panel), initial frequency (bottom left panel), and S/N (bottom right panel) distributions of the detectable mock BBHs by AMIGO for the EMBS and dynamical models, respectively. Light blue shaded histograms, blue dotted, and solid histograms show the corresponding results from the EMBS model as explained in Fig.~\ref{fig:f3}, and the light purple shaded histogram, purple dotted, and solid histograms show the results from the dynamical model.}
\label{fig:f4}
\end{figure*}

Figure~\ref{fig:f3} illustrates the redshift (top left panel), chirp mass (top right panel), initial frequency (bottom left panel), and S/N (bottom right panel) distributions of the detectable BBHs by AMIGO ($\varrho_{\rm AMIGO} \geq 8$) over a continuous observation period of $4$ yr. The mock BBH sample is from the $50$th among the $100$ realizations generated from the GWTC-3 and hybrid models, respectively,  with the local merger rate density of $19.1 \gpcyr$. Figure~\ref{fig:f4} shows the same distributions obtained from the EMBS and dynamical models, respectively. As seen from these two figures, the most detectable sources are nearby systems with $z < 0.5$ and heavy ones with $\mathcal{M}_{\rm c}$ in the range from $20$ to $50\,\rm M_{\odot}$. Their initial frequencies mostly lie in the range from $0.01$ to $0.1$\,Hz as the middle-frequency GW detectors achieve the best sensitivity in this frequency range. The distributions of these parameters for the detectable BBHs resulting from different models do have some differences but mainly contributed by the variances among different realizations and the uncertainty in the local merger rate density. For simplicity, we illustrate the model results below only for the GWTC-$3$ model with the local merger rate density fixed at $19.1$\,Gpc$^{-3}$\,yr$^{-1}$.

\begin{figure}
\centering
\includegraphics[width=\columnwidth]{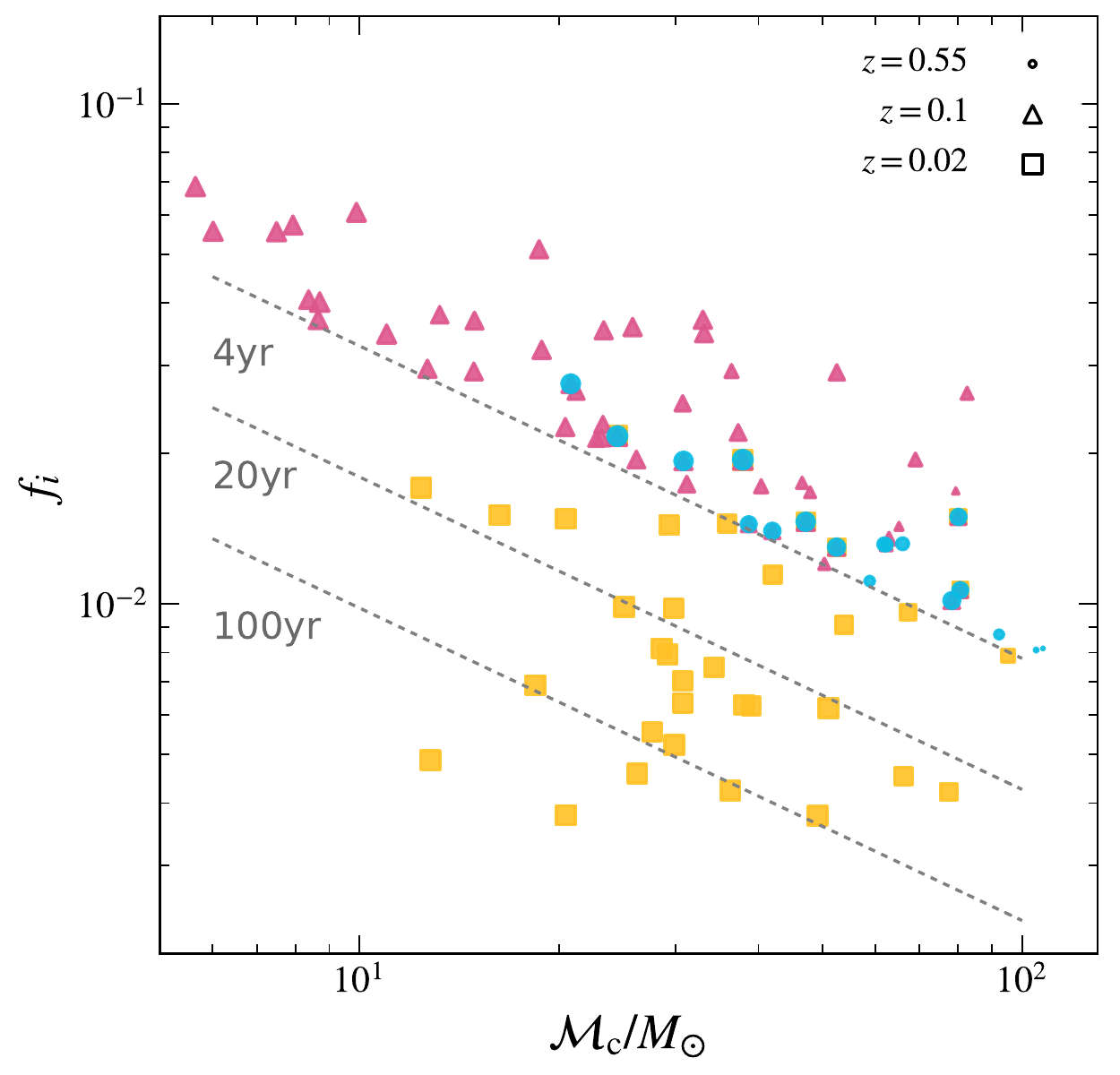}
\caption{Redshifted chirp mass and observed GW frequency at the beginning of the observation of mock BBHs detected by the joint observation of LT-AMIGO.
Cyan circles represent the multiband BBHs detected by both LT and AMIGO (i.e. $\varrho_{\rm LT} \geq 5$ and $\varrho_{\rm AMI} \geq 5$). Magenta triangles represent the BBHs detected by AMIGO alone (i.e., $\varrho_{\rm AMI} \geq 8$). Yellow squares represent the BBHs detected by LT alone (i.e., $\varrho_{\rm LT} \geq 8$). The redshift of the BBHs are illustrated by the size of each symbol (see the labels in this figure). The dotted lines show the BBHs with the same lifetime (e.g., $4$, $20$, and $100$\,yr) but different $\mathcal{M}_{\rm c}$ and $f_{\rm i}$.}
\label{fig:f5}
\end{figure}

\begin{figure}
\centering
\includegraphics[width=\columnwidth]{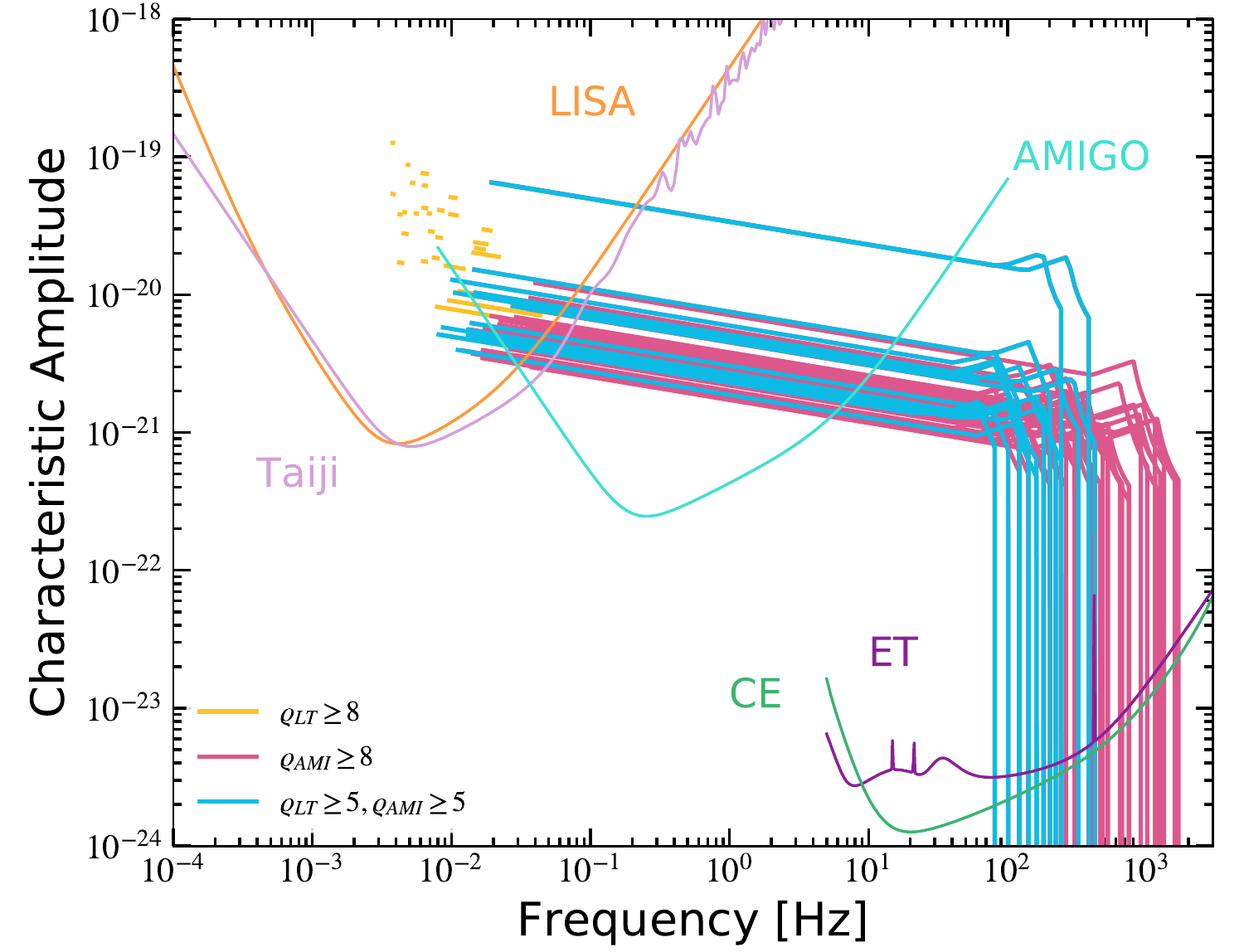}
\caption{Evolution tracks of characteristic amplitudes for the mock BBHs which are detectable by LT-AMIGO (i.e., $\varrho_{\rm LT} \geq 5$ and $\varrho_{\rm AMI} \geq 5$, blue lines), by  LT  (i.e., $\varrho_{\rm LT} \geq 8$, yellow lines), and by  AMIGO (i.e., $\varrho_{\rm AMI} \geq 8$, magenta lines), respectively. All the BBHs plotted here are observed for a continuous period of $4$ yr. Orange, pink, sky blue, green, and purple curves represent the sensitivity curves for LISA, Taiji, AMIGO, CE, and ET, respectively. 
}
\label{fig:f6}
\end{figure}

Figure~\ref{fig:f5} shows the chirp mass, the frequency at the beginning of the observation 
of each detectable BBH  from the 50th among the 100 realizations generated from the GWTC-3 model regarding to LT, AMIGO, and LT-AMIGO joint observations, respectively. Figure~\ref{fig:f6} illustrates the characteristic strain evolution of the multiband BBHs shown in Figure~\ref{fig:f5}. There are $18$ multiband BBHs (i.e., $\varrho_{\rm LT} \geq 5$ and $\varrho_{\rm AMI} \geq 5$, cyan circles in Fig.~\ref{fig:f5} and cyan lines in Fig.~\ref{fig:f6}), which can be detected by LT first, then detected by AMIGO in the middle-frequency, and finally detected by CE/ET at the merging and ringdown phases. LT can detect $35$ BBHs with $\varrho_{\rm LT} \geq 8$ (yellow squares in Fig.~\ref{fig:f5} and yellow lines in Fig.~\ref{fig:f6}), of which $6$ BBHs are also multiband ones. About $80\%$ of the BBHs detectable by LT cannot evolve to the middle-frequency and merge within a continuous observation period of $4$ yr. AMIGO may detect $51$ BBHs with $\varrho_{\rm AMI} \geq 8$ (magenta squares in Fig.~\ref{fig:f5} and magenta lines in Fig.~\ref{fig:f6}), of which $12$ BBHs are multiband ones. All of these $51$ BBHs can merge within a continuous observation period of $4$ yr and can be detected by the high-frequency GW detectors, but only $\sim 18\%$ of them has $\varrho_{\rm LT}$ larger than $5$. The number of multiband BBHs is smaller than those detectable by either LT or AMIGO alone because of the following reasons. A large fraction of BBHs detected by LT cannot evolve to AMIGO band within $4$ yr, and most of the BBHs detected by AMIGO do not exceed the detection threshold in the LT band.
For the particular case considered here, the BBHs detectable in the middle-frequency band (magenta triangles in Fig.~\ref{fig:f5}) can always evolve into the high-frequency band within the $4$ yr observation period, while not all those detectable in the low-frequency band can do. 

\section{Precision of Parameter Estimations}
\label{sec:PE}

We analyse the parameter estimation accuracy of these sources by GW detectors in the different frequency bands and the networks formed by these detectors, and figure out how the estimations of the source parameters being improved by multiband GW observations in this section. We adopt the commonly used FIM method, which is valid in the limit of high S/N, as many other studies \citep{Cutler1994, Grimm2020, LiuShao2020, ChenLu2021}. We note here that several assumptions made in the FIM method may not be valid in some cases, but it can always provide a quick rough estimations about the precision of the GW source parameter estimates for those sources with high S/N. More discussions can be found in Section~\ref{sec:dis_con}. The Fisher matrix can be calculated by
\be
\label{eq:Gamma}
\Gamma_{ab} =  \left(\left. \frac{\partial h}{\partial \Xi^{a}} \right| \frac{\partial h}{\partial \Xi^{b}}\right),
\ee
where $\Xi$ represents the stellar BBH parameters. In our calculation, we consider seven parameters: $d_{\rm L}$, $\mc$, $\eta$, $t_{\rm c}$, $\phi_{\rm c}$, $\theta_{\rm S}$, and $\phi_{\rm S}$. For multidetector networks, the Fisher matrix is as follows
\begin{eqnarray}
\label{eq:multiGamma}
\Gamma_{a b} & = & \left(\left.\frac{\partial h}{\partial \boldsymbol{\Xi}^{a}} \right| \frac{\partial h}{\partial \boldsymbol{\Xi}^{b}}\right) \nonumber \\
& = & \sum_{j=1}^{n} 2\int_{f_{\rm i}}^{f_{\rm f}} \frac{\frac{\partial \tilde{h}_j^{*}(f)}{\partial \boldsymbol{\Xi}^{a}}\frac{\partial \tilde{h}_j(f)}{\partial \boldsymbol{\Xi}^{b}}+\frac{\partial \tilde{h}_j^{*}(f)}{\partial \boldsymbol{\Xi}^{b}}\frac{\partial \tilde{h}_j(f)}{\partial \boldsymbol{\Xi}^{a}}}{S_{{\rm n},j}(f)} df,
\end{eqnarray}
where $j$ and $n$ have the same meaning as in Equation~\eqref{eq:snrmulti}. Taking the LISA-Taiji network as an example, $j = 1$ refers to one of the two Michelson interferometers of LISA with $\alpha_0 = 0$ and $\Phi_0 = 0$, while $j = 2$ refers to the other interferometer of LISA with $\alpha_0 = \pi/4$ and $\Phi_0 = 0$. As for Taiji, $\alpha_0$ equals to $\Delta\alpha$ and $\Delta\alpha+\pi/4$ for two interferometers respectively, and $\Phi_0 = 2\pi/9$ since LISA is behind the Earth by about $20^{\circ}$ in a heliocentric orbit while Taiji is ahead the Earth by about $20^{\circ}$ in a heliocentric orbit. We adopt the choice that AMIGO is behind the Earth by about $2^{\circ}$ in a heliocentric orbit \citep{Ni2022}, thus $\Phi_0 = \pi/10$, $\alpha_0 = \Delta\alpha_1$ and $\alpha_0 = \Delta\alpha_1+\pi/4$ for the two interferometers, respectively. Since the initial orientation between different GW detectors will not affect the result much, we assume $\Delta\alpha = \pi/2$ and $\Delta\alpha_1 = \pi/3$ \citep{ChenLu2021}. We also calculate the Fisher matrix regarding the ground-based high-frequency GW detectors, like CE and ET, about those  detectable BBHs by the low-frequency or middle-frequency GW detectors. Since we generate the locations of mock BBHs in the ecliptic coordinate frame, we transform ecliptic coordinates into ground-based GW detector frame the same as the work of \citet{Jaranowski1998}. We consider the effect of Earth rotation like the work of \citet{Zhaowen2018} as well. The ground-based detector sites influence the detector's antenna pattern function through the detector's latitude $\varphi$, longitude $\lambda$, the angle between the detector arm and local geographical direction $\gamma$ and the angle between the interferometer arms $\zeta$. These parameters are listed in Table~\ref{tab:site} similar as those in the work of \citet{Zhaowen2018}. We assume that CE consists two interferometers locating at the same places as two LIGO interferometers. Given the Fisher matrix, the covariance matrix can be calculated by
\be
\Sigma = \left\langle\delta \Xi^{a} \delta \Xi^{b}\right\rangle = \left(\Gamma^{-1}\right)^{a b}.
\ee
The uncertainties of BBHs parameters estimates is then given by 
\be
\Delta \boldsymbol{\Xi}^{a}=\sqrt{\left(\Gamma^{-1}\right)^{a a}}.
\ee
Note that the angular resolution $\Delta\Omega$ at given confidence level ($\rm{X}\%$) is defined as \citep{Cutler1994,Barack2004,WenChen2010}
%
\be
\Delta\Omega_{{\rm X}\%}=-2\pi|\sin\theta_{\rm S}|\sqrt{(\Delta\theta_{\rm S}\Delta\phi_{\rm S})^2-\left\langle\Delta\theta_{\rm S}\Delta\phi_{\rm S}\right\rangle^2}\ln(1-{\rm X}/100),
\ee
%
where $\Delta\theta_{\rm S}$, $\Delta\phi_{\rm S}$, $\left\langle\Delta\theta_{\rm S}\Delta\phi_{\rm S}\right\rangle$ are obtained from the covariance matrix. We adopt $90\%$ credible angular resolution ($\Delta\Omega_{90\%}$) as the sky localization area throughout this paper.

\begin{table}
\centering
\caption{
Location sites of ground-based high frequency GW detectors.
}
\resizebox{1\linewidth}{!}{
\begin{tabular}{c|c|c|c|c} %
\hline		
\hline
 & $\varphi$ & $\lambda$ & $\gamma$ & $\zeta$ \\ \hline
Einstein Telescope & $43.54^{\circ}$ & $10.42^{\circ}$ & $19.48^{\circ}$ & $60^{\circ}$ \\ \hline
Cosmic Explorer in Hanford & $46.45^{\circ}$ & $-119.41^{\circ}$ & $90.95^{\circ}$ & $90^{\circ}$ \\ \hline
Cosmic Explorer in Livingston & $30.54^{\circ}$ & $-90.77^{\circ}$ & $162.15^{\circ}$ & $90^{\circ}$ \\ \hline
\end{tabular}
}
\begin{flushleft}
\footnotesize{Note: first and second columns denote the latitude and longitude of the detector. Third column lists the angle that is measured counter-clockwise from East of the local geographical direction to the base vector of the detector's arm. Last column represents the angle between the interferometer arms.}
\end{flushleft}
\label{tab:site}
\end{table}

\begin{figure}
\centering
\includegraphics[width=\columnwidth]{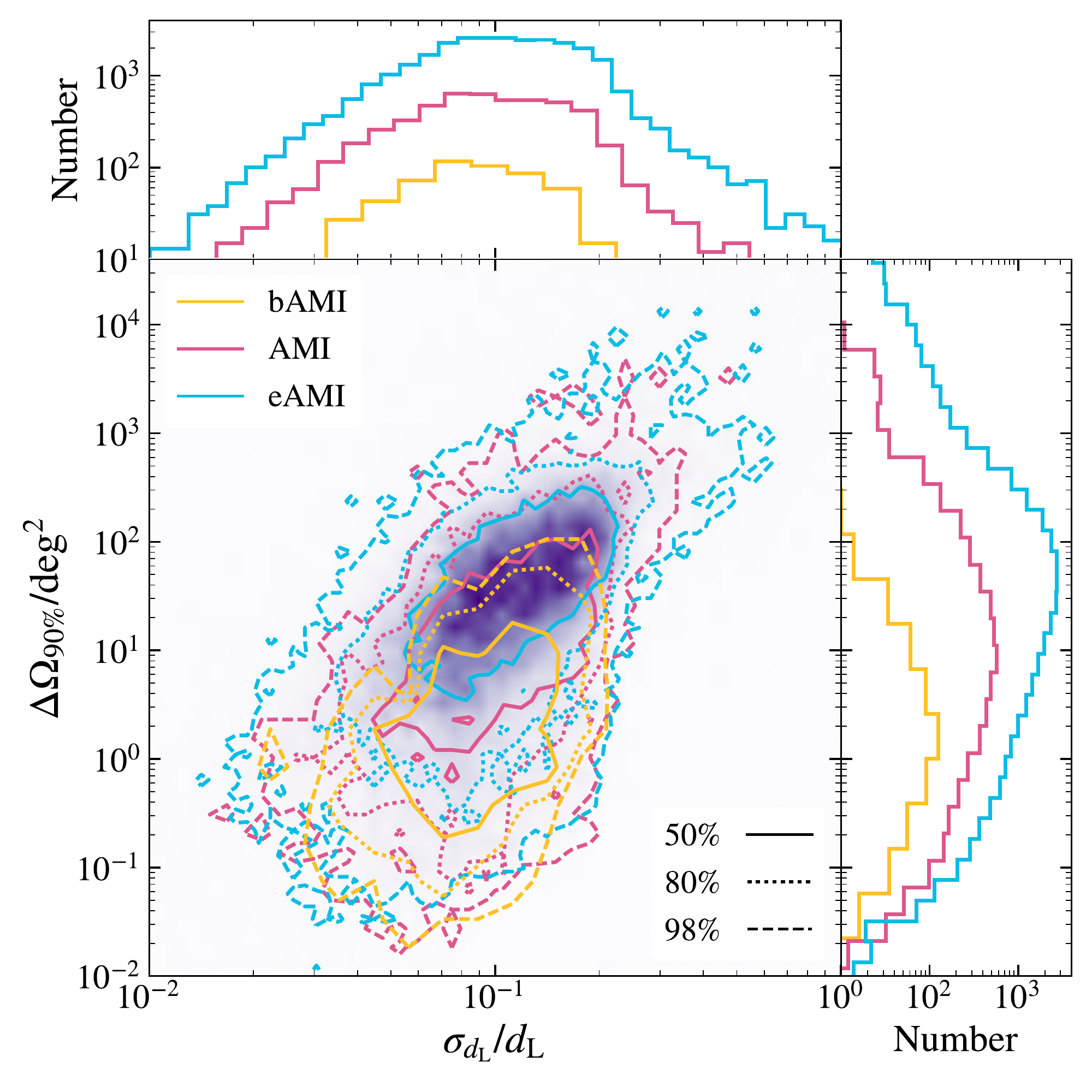}
\caption{Contours for the errors of sky localization and the relative error of luminosity distance measurements for those mock BBHs detectable for bAMIGO (yellow), AMIGO (magenta), and eAMIGO (cyan) averaged over all the $100$ realizations generated from the GWTC-3 model with the local merger rate density of $19.1\gpcyr$. The mock BBHs within the area labelled by solid (dotted/dashed) contours account for $50\%$ ($80\%$/$98\%$) of all the detectable sources. The small panels above/right to the main panel show the histograms of the relative errors of the luminosity distance measurements/the sky localization precision for those detectable BBHs by bAMIGO (yellow), AMIGO (magenta), and eAMIGO (cyan), respectively.
}
\label{fig:f7}
\end{figure}

\begin{figure}
\centering
\includegraphics[width=\columnwidth]{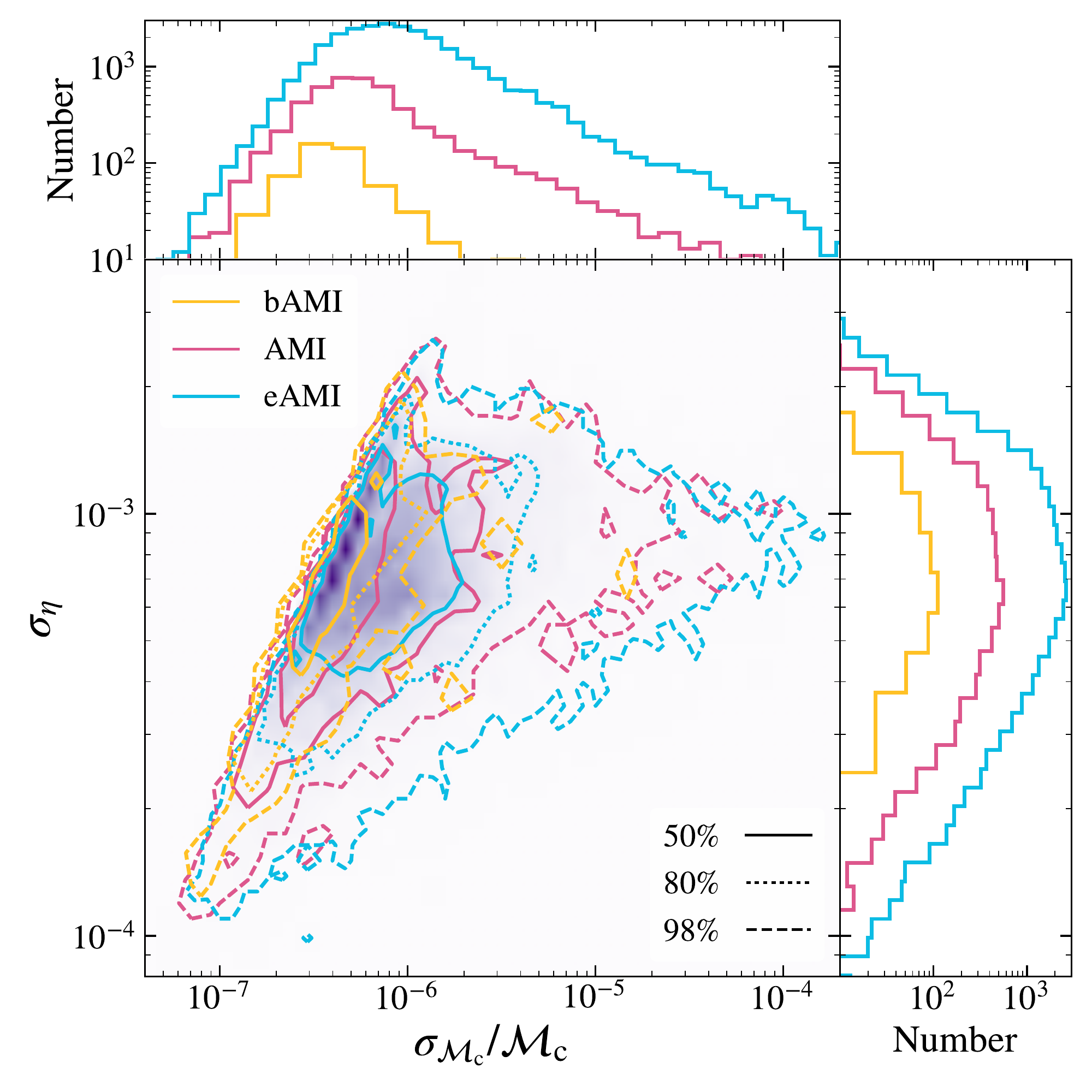}
\caption{
Legend is similar to that for Fig.~\ref{fig:f7}, except that this figure is for the errors of the symmetric mass ratios ($\sigma_\eta$) and the relative errors of redshifted chirp mass ($\sigma_{\mathcal{M}_{\rm c}}/\mathcal{M}_{\rm c}$) of mock BBHs.
}
\label{fig:f8}
\end{figure}

Figures~\ref{fig:f7} and \ref{fig:f8} show the probability distributions of the sky localization area,  the relative errors of distances, redshifted chirp masses, and symmetric mass ratios of those BBHs detected by bAMIGO, AMIGO, and eAMIGO among all $100$ realizations generated from the GWTC-$3$ model with $\Rmrgbh = 19.1\gpcyr$. For these $100$ realizations, there are totally $\sim 550$ BBHs with $\varrho_{\rm bAMI} \geq 8$, $\sim 5100$ BBHs with $\varrho_{\rm AMI} \geq 8$, and $\sim 29000$ BBHs with $\varrho_{\rm eAMI} \geq 8$. bAMIGO, AMIGO, and eAMIGO may localize around $98\%$ of their detectable sources in the sky areas with $\Delta \Omega_{90\%} \sim$ ($0.02$\,deg$^2$, 
$100$\,deg$^2$), ($0.02$\,deg$^2$, $5000$\,deg$^2$) and ($ 0.01$\,deg$^{2}$,$10^{4}$\,deg$^{2}$)
, respectively. The distributions of $\Delta \Omega_{90\%}$ for bAMIGO, AMIGO, and eAMIGO sources peak around 
$2$\,deg$^2$, $10$\,deg$^2$, and $50$\,deg$^2$,
respectively. The relative errors of the luminosity distance measurements for $98\%$ of the detectable BBHs by bAMIGO, AMIGO, and eAMIGO are in the range of 
$\sim (0.02, 0.2)$, $(0.01,0.5)$, and $(0.01,0.6)$
, respectively, and their distributions all peak around $0.1$.
eAMIGO and AMIGO can detect more distant sources, which are relatively hard to be localized and have less accurate parameter estimations than those close sources detected by bAMIGO. 
Similarly, Figure~\ref{fig:f8} shows the estimation precision of the redshifted chirp mass ($\sigma_{\mathcal{M}_{\rm c}}/\mathcal{M}_{\rm c}$) and symmetric mass ratio ($\sigma_{\eta}$) for BBHs detected by bAMIGO, AMIGO, and eAMIGO. The distributions of $\sigma_{\mathcal{M}_{\rm c}}/\mathcal{M}_{\rm c}$ for bAMIGO, AMIGO, and eAMIGO centre around $4\times10^{-7}$, $6\times10^{-7}$, and $8\times10^{-7}$
, respectively; the distributions of $\sigma_{\eta}$ for bAMIGO, AMIGO, and eAMIGO all centre around $10^{-3}$, respectively.
The two distributions for eAMIGO have the largest scatters, which are due to that (1) eAMIGO has the highest sensitivity and can detect those close BBHs with higher S/Ns, and (2) eAMIGO detects more distant sources, compared with AMIGO and bAMIGO. For those sources that can be detected by bAMIGO, eAMIGO observations can slightly improve the precision of measurements of $\mathcal{M}_{\rm c}$ and $\eta$ by a factor of several compared with that by bAMIGO.
               
\begin{figure*}
\centering
\includegraphics[scale=0.6]{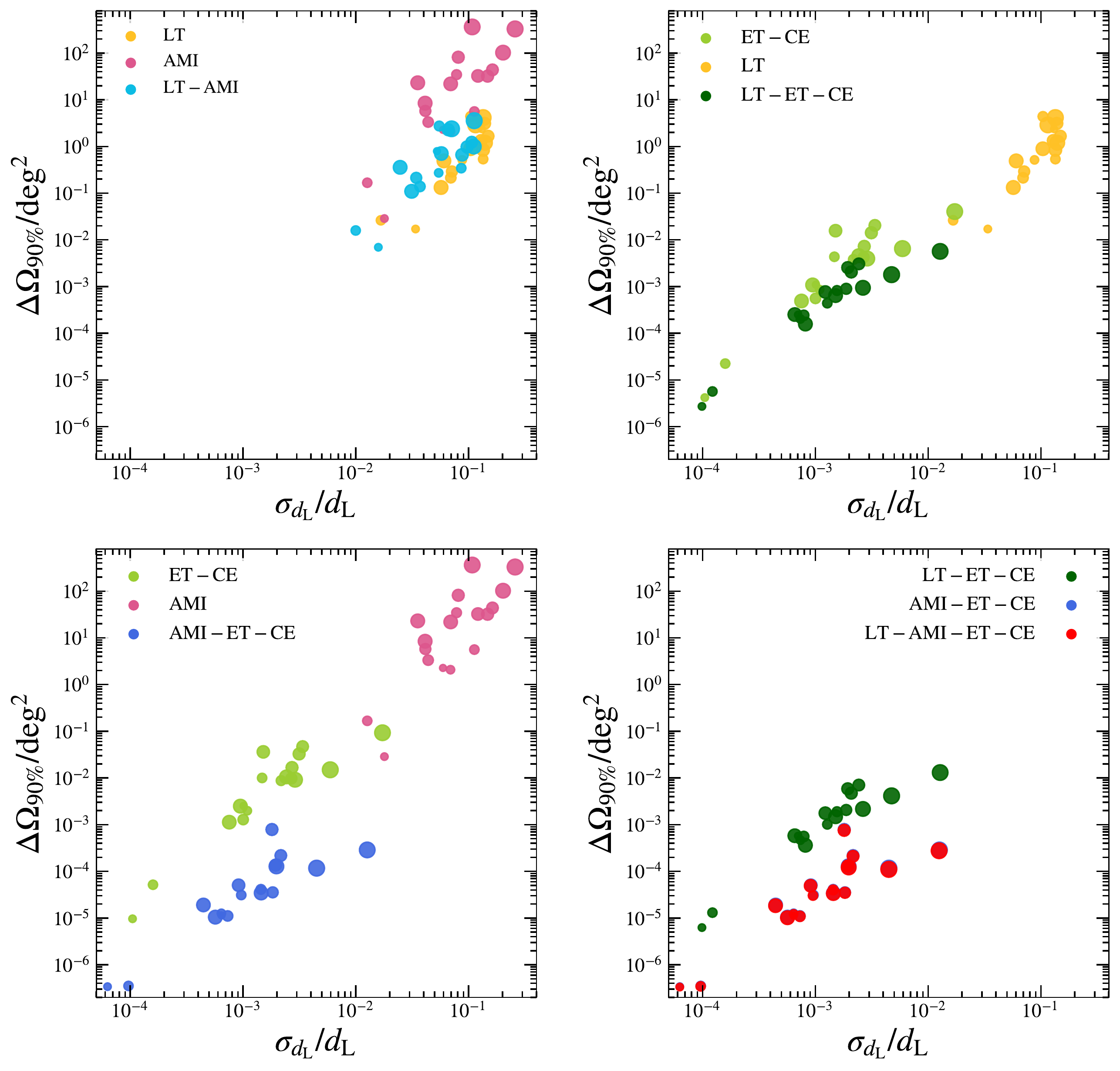}
\caption{
The estimated sky localization errors and relative errors of luminosity distance measurements for the multiband BBHs from the $50$th realization ranking by detectable number among the $100$ realizations randomly generated from the GWTC-$3$ model with $\Rmrgbh = 19.1\gpcyr$, detected by the multiband GW detectors and the joint observations by them. Top left panel shows the results for multiband BBHs detected by LT, AMIGO, and LT-AMIGO, respectively. Top right panel shows the results for multiband BBHs detected by LT, ET-CE, and LT-ET-CE, respectively. Bottom left panel shows the results for multiband BBHs detected by ET-CE, AMIGO, and AMIGO-ET-CE, respectively. Bottom right panel shows the results for multiband BBHs detected by LT-AMIGO-ET-CE. Different colours represent the results on parameter estimation errors for various multiband GW detectors and their joint observations as labelled by texts in each panel, i.e., yellow, magenta, cyan, light green, dark green, blue, and red colours represent those by LT, AMIGO, LT-AMIGO, ET-CE, LT-ET-CE, AMIGO-ET-CE, and LT-AMIGO-ET-CE, respectively.
}
\label{fig:f9}
\end{figure*}

\begin{figure*}
\centering
\includegraphics[scale=0.65]{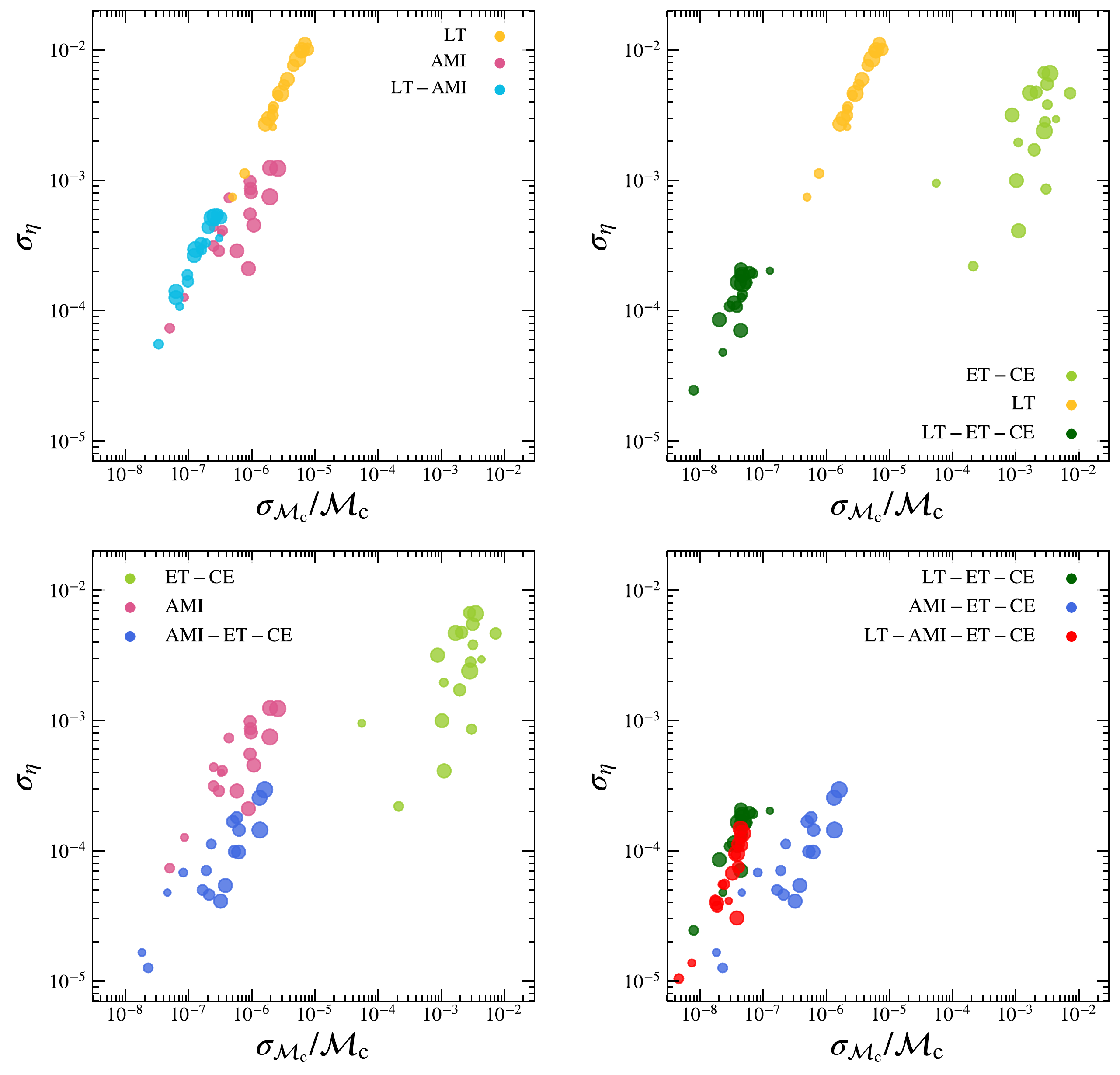}
\caption{The estimated relative errors of the redshifted chirp masses and the symmetric mass ratios for multiband BBHs detected by the multiband GW detectors and the networks combined by them. Symbols and colours have the same meanings as those in Fig.~\ref{fig:f9}.}
\label{fig:f10}
\end{figure*}

\begin{figure*}
    \centering
    \includegraphics[scale=0.7]{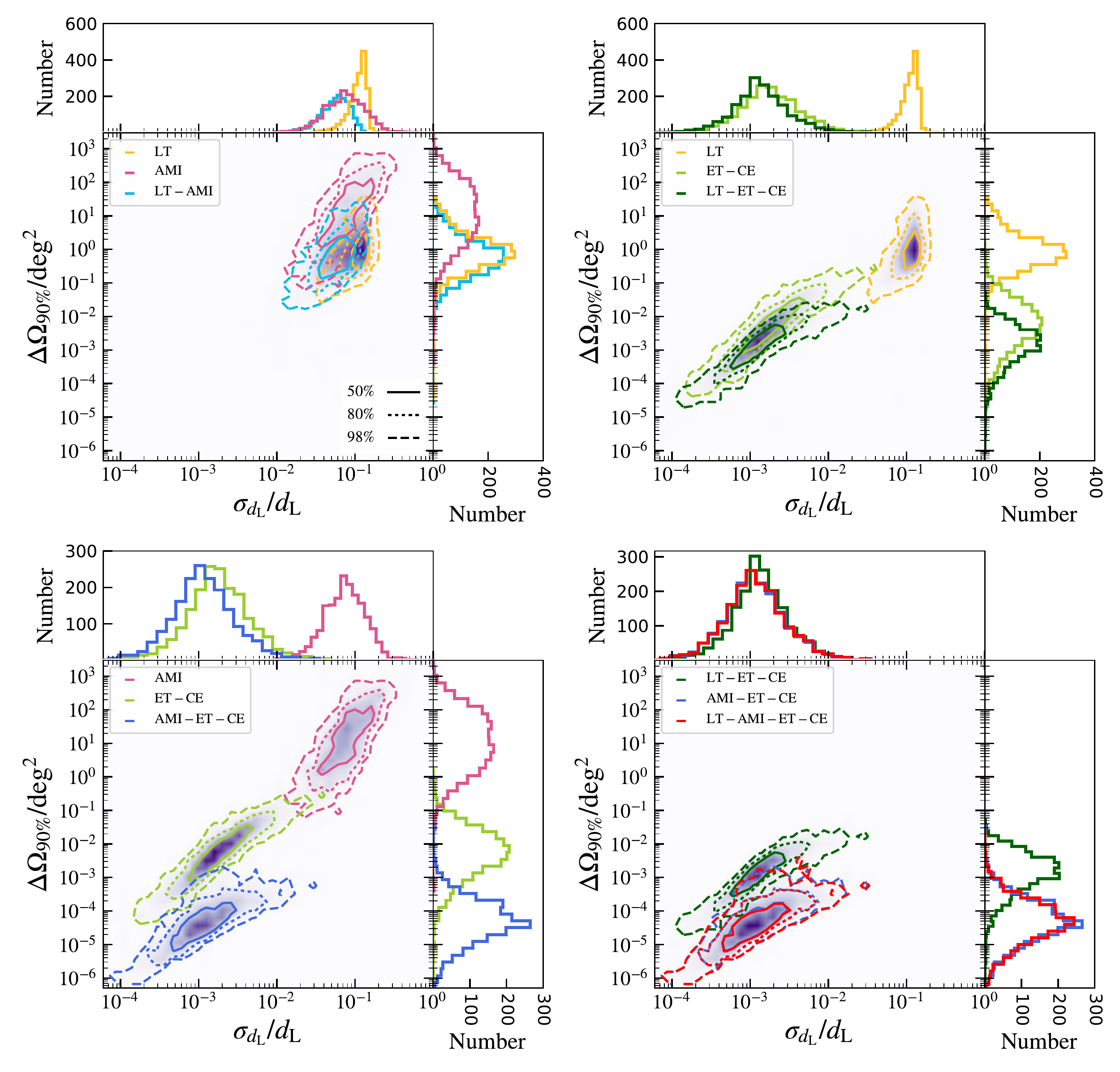}
    \caption{Contours of the estimated sky localization errors and relative errors of luminosity distance measurements for multiband BBHs from all the $100$ realizations randomly generated from the GWTC-$3$ model with the local merger rate density of $\Rmrgbh = 19.1\gpcyr$. Dashed (dotted/solid) contours in each panel represent the region where contains $98\%$ ($80\%$/$50\%$) of all the multiband BBHs. Top left panel shows the results for the multiband BBHs detected by LT (yellow), AMIGO (magenta), and LT-AMIGO (cyan), respectively. Top right panel shows the results for the multiband BBHs detected by LT (yellow),  ET-CE (light green), and LT-ET-CE (dark green), respectively. Bottom left panel shows the results for the multiband BBHs detected by ET-CE (light green), AMIGO (magenta), and AMIGO-ET-CE (blue), respectively. Bottom right panel shows the results for the multiband BBHs detected by LT-ET-CE (dark green), AMIGO-ET-CE (blue), and LT-AMIGO-ET-CE (red), respectively.}
    \label{fig:f11}
\end{figure*}

\begin{figure*}
    \centering
    \includegraphics[scale=0.7]{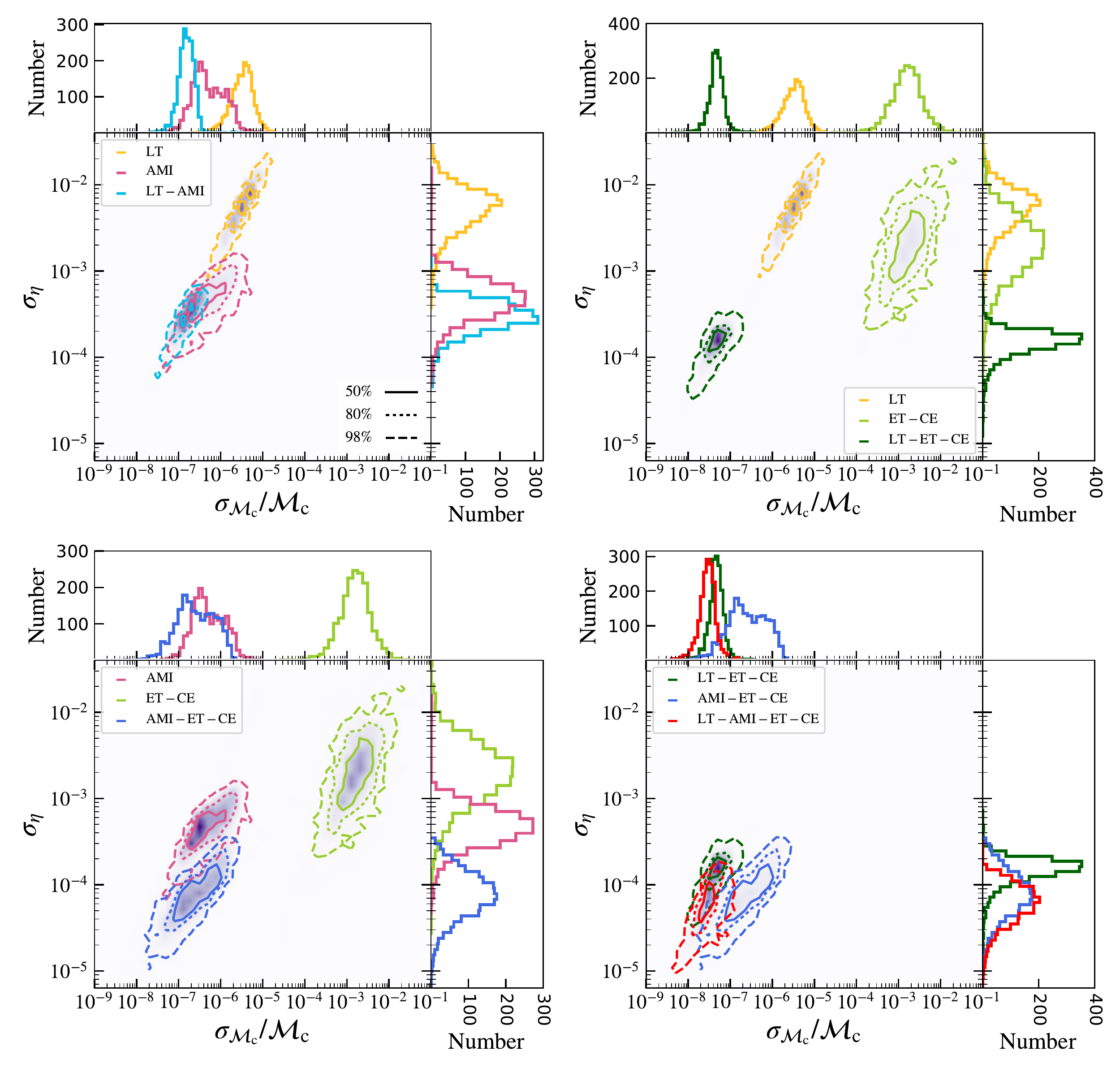}
    \caption{Legend is similar to that for Fig.~\ref{fig:f11}, except that for the relative measurement errors of the redshifted chirp mass $\mc$ and symmetric mass ratio $\eta$ for multiband BBHs.
    }
    \label{fig:f12}
\end{figure*}

Figures~\ref{fig:f9} and \ref{fig:f10} illustrate the effects of multiband joint observations on the localization and physical parameter measurements of BBHs. These mock BBHs are adopted from the $50$th realization, among $100$ realizations generated from the GWTC-$3$ model with $\Rmrgbh = 19.1\gpcyr$, ranking by the detection numbers expected from the LT-AMIGO joint observations. There are $18$ BBHs meet our definition for detectable multiband BBHs in this realization. As the number of mock detectable BBHs from a single realization is small, the distributions shown in Figures~\ref{fig:f9} and \ref{fig:f10} may deviate from the actual distributions because of large sample variance. We also plot those detectable BBHs by contours from all the $100$ realizations in Figures~\ref{fig:f11} and \ref{fig:f12}, with which the distributions can be close to the actual ones and may be affected little by sample variance.

The top left panels of Figures~\ref{fig:f9} and \ref{fig:f11} show the localization ability and luminosity measurement precision for the detectable multiband BBHs from the $50$th realization and those from all the $100$ realizations by LT, AMIGO, and LT-AMIGO, respectively. For the $50$th realization (Fig.~\ref{fig:f9}), LT may localize the BBHs in sky areas of $\sim 0.02$ to $2\,\rm deg^2$, about one order of magnitude better than AMIGO does (with $\Delta \Omega_{90\%} \sim (0.03,300)$\,deg$^2$). The estimated relative errors of luminosity distances by LT and AMIGO both centre around $0.1$. For all the $100$ realizations (Fig.~\ref{fig:f11}), LT localize $98\%$ of multiband BBHs in sky areas of $\sim 0.03$ to $40\,\rm deg^2$, while AMIGO may localize them in sky areas of $0.07-700\,\rm deg^2$.
The median values of the distributions of $\Delta \Omega_{90\%}$ and $\sigma_{d_{\rm L}}/d_{\rm L}$ by LT/AMIGO/LT-AMIGO are $0.82$/$1.1$/$0.54$\,deg$^2$ and $0.11$/$0.075$/$0.058$ (see Tab.~\ref{tab:impro}), and their corresponding $68\%$ confidence intervals are 
$0.083$-$8.1$/$1.1$-$110$/$0.055$-$5.4$\,deg$^2$ and $0.012$-$1.1$/$7.6\times10^{-3}$-$0.74$/$5.8$$\times$$10^{-3}$-$0.57$.
Taking the ratios of the median values of each parameter estimation uncertainty distributions from different types of GW detectors as the improvement, LT-AMIGO may improve the localization by a factor $1.5$ or $20$ compared with the observations by LT or AMIGO only. Since LT has longer baseline than AMIGO and the sources spend more time during the LT frequency band, the localization ability is dominated by LT rather than AMIGO when combining LT and AMIGO together. Similarly the improvement on the precision of luminosity distance measurements by LT-AMIGO can be a factor $2$ or $1.3$ compared with those by LT or AMIGO only. The differences between the precision on luminosity distance measured by LT observations and that by AMIGO observations are not significant, because $\sigma_{d_{\rm L}}/d_{\rm L}$ mainly depends on the S/N of a source and the S/N obtained by LT observations is not significant different from that by AMIGO observations for an individual source.

The top left panel of Figure~\ref{fig:f10} shows that the AMIGO observations give more accurate estimations on $\mathcal{M}_{\rm c}$ and $\eta$ than those by LT observations for the $18$ multiband BBHs. The top left panel of Figure~\ref{fig:f12} shows that AMIGO may give the measurements of $\mathcal{M}_{\rm c}$ with relative errors ranging from $5\times10^{-8}$ to 
$5\times10^{-6}$ and $\eta$ with errors between 
$8\times10^{-5}$ and $10^{-3}$ for $98\%$ of the multiband BBHs from all $100$ realizations, while LT observations may measure them with errors about several times to one order of magnitude larger than those by AMIGO, respectively. The median values of $\sigma_{\mc}/\mc$ and $\sigma_{\eta}$ distributions for multiband BBHs from all the $100$ realizations given by LT/AMIGO/LT-AMIGO are $3.4$$\times$$10^{-6}$/$4.5$$\times$$10^{-7}$/$1.5$$\times$$10^{-7}$ and $5.5$$\times$$10^{-3}$/$4.7$$\times$$10^{-4}$/$3.0$$\times$$10^{-4}$.
The $68\%$ confidence intervals for these distributions are 
$3.4$$\times$$10^{-7}$-$3.3$$\times$$10^{-5}$/$4.6$$\times$$10^{-8}$-$4.4$$\times$$10^{-6}$/$1.6$$\times$$10^{-8}$-$1.5$$\times$$10^{-6}$ and $5.6$$\times$$10^{-4}$-$0.054$/$4.8$$\times$$10^{-5}$-$4.7$$\times$$10^{-3}$/$3.0$$\times$$10^{-5}$-$2.9$$\times$$10^{-3}$
(Tab.~\ref{tab:impro}). LT-AMIGO may improve the estimation accuracy of $\mathcal{M}_{\rm c}$ and $\eta$ by about a factor $21$ and $18$ (or about $3$ and $1.6$) compared with those by LT (or AMIGO) alone. AMIGO is able to measure the redshifted chirp mass and symmetric mass ratio more accurately because that the frequency evolution of BBHs is faster in the AMIGO band than in the low-frequency band.

The top right panel of Figure~\ref{fig:f9} illustrates the effects on localization and measurement accuracy of luminosity distances by the joint observations of the low- and high-frequency GW detectors for the multiband BBHs in the $50$th realization. ET-CE may localize them in the sky areas of $4\times10^{-6}$ to $0.04\,\rm{deg^2}$
and measure the luminosity distances with relative errors ranging from $10^{-4}$ to $0.02$, which are significantly better than the LT results since those BBHs possess high S/Ns in the high-frequency band. If considering sources from all the $100$ realizations (top right panel of Fig.~\ref{fig:f11}), ET-CE/LT-ET-CE may localize $98\%$ of those multiband BBHs in the sky areas of 
$4$$\times$$10^{-5}$-$0.3$\,deg$^2$/$2$$\times$$10^{-5}$-$0.02$\,deg$^2$ and measure luminosity distance with relative errors in the range of $2$$\times$$10^{-4}$-$0.04$/$10^{-4}$
-$0.02$. The median values for the distributions of $\Delta \Omega_{90\%}$ and $\sigma_{d_{\rm L}/d_{\rm L}}$ by ET-CE/LT-ET-CE are 
$5.7\times10^{-3}$/$1.5\times10^{-3}$\,deg$^2$ and $1.8$$\times$$10^{-3}$/$1.3$$\times$$10^{-3}$,
with $68\%$ confidence intervals as 
$5.8$$\times$$10^{-4}$-$0.057$\,deg$^2$/$1.5$$\times$$10^{-4}$-$0.015$\,deg$^2$ and $1.8$$\times$$10^{-4}$-$0.017$/$1.3$$\times$$10^{-4}$-$0.013$, respectively. Compared to the results of LT or ET-CE only, LT-ET-CE will improve the localization by a factor $\sim 550$ or $4$
and the luminosity distance measurement accuracy by a factor $\sim 84$ or $1.3$.

The top right panel of Figure~\ref{fig:f10} illustrates the effects on the measurement accuracy of $\mathcal{M}_{\rm c}$ and $\eta$ by LT-ET-CE joint observations for those $18$ BBHs in the $50$th realization. ET-CE gives larger relative errors of $\mathcal{M}_{\rm c}$, ranging from 
$6\times10^{-5}$ to $7\times10^{-3}$,
than LT does, while they give comparable estimations about the errors of $\eta$, ranging from 
$2\times10^{-4}$ to $7\times10^{-3}$.
The top right panel of Figure~\ref{fig:f12} shows that LT-ET-CE may measure $\mc$ for $98\%$ of multiband BBHs from all the $100$ realizations with relative errors in the range of 
$10^{-8}$-$2$$\times$$10^{-7}$,
and $\eta$ with errors between $3$$\times$$10^{-5}$ and $3$$\times$$10^{-4}$.
The distribution of $\sigma_{\mc}/{\mc}$ obtained from the joint observations by LT-ET-CE has a median value of $4.7$$\times$$10^{-8}$ and its $68\%$ confidence interval is $4.7$$\times$$10^{-9}$-$4.6$$\times$$10^{-7}$, and the distribution of $\sigma_\eta$ has a median value of $1.5$$\times$$10^{-4}$ and its $68\%$ confidence interval is $1.5$$\times$$10^{-5}$-$1.5$$\times$$10^{-3}$.
This indicates that the measurement precision of $\mathcal{M}_{\rm c}$ and $\eta$ by the LT-ET-CE joint observations is improved by a factor 
$\sim 72$ (or $\sim 3\times10^{4}$) and $\sim 37$ (or $\sim 13$), compared with those by LT (or ET-CE) observations only, respectively.


\begin{table*}
\centering
\caption{
{Median values and $68$ percent confidence intervals of the distributions of parameter estimation uncertainties for multiband BBHs}
}
\label{tab:impro}
\resizebox{1\linewidth}{!}{
\begin{tabular}{c|c|c|c|c|c|c|c|c|c|c|c|c} %
\hline		
\hline
\multirow{2}{*}{GW Detector}  & \multicolumn{3}{c}{$\Delta \Omega_{90\%}$}  & \multicolumn{3}{c}{$\sigma_{d_{\rm L}}/d_{\rm L}$}  & \multicolumn{3}{c}{$\sigma_{\mc}/\mc$}  & \multicolumn{3}{c}{$\sigma_\eta$}  \\  
& ${\rm low}$ & ${\rm median}$ & ${\rm high}$ & ${\rm low}$ & ${\rm median}$ & ${\rm high}$ & ${\rm low}$ & ${\rm median}$ & ${\rm high}$ & ${\rm low}$ & ${\rm median}$ & ${\rm high}$ \\ \hline
 %
  {LT} & $8.3$$\times$$10^{-2}$ & $8.2$$\times$$10^{-1}$ & $8.1$$\times$$10^{0}$ & $1.2$$\times$$10^{-2}$ & $1.1$$\times$$10^{-1}$ & $1.1$$\times$$10^{0}$ & $3.4\times10^{-7}$ & $3.4\times10^{-6}$ & $3.3\times10^{-5}$ & $5.6\times10^{-4}$ & $5.5\times10^{-3}$ & $5.4\times10^{-2}$ \\ \hline
 %
 {AMIGO} & $1.1\times10^{0}$ & $1.1\times10^{1}$ & $1.1\times10^{2}$ & $7.6\times10^{-3}$ & $7.5\times10^{-2}$ & $7.4\times10^{-1}$ & $4.6\times10^{-8}$ & $4.5\times10^{-7}$ & $4.4\times10^{-6}$ & $4.8\times10^{-5}$ & $4.7\times10^{-4}$ & $4.7\times10^{-3}$ \\ \hline
 %
 {ET-CE} & $5.8\times10^{-4}$ & $5.7\times10^{-3}$ & $5.7\times10^{-2}$ & $1.8\times10^{-4}$ & $1.8\times10^{-3}$ & $1.7\times10^{-2}$ & $1.6\times10^{-4}$ & $1.6\times10^{-3}$ & $1.6\times10^{-2}$ & $2.0\times10^{-4}$ & $2.0\times10^{-3}$ & $1.9\times10^{-2}$ \\ \hline

 %
 {LT-AMIGO} & $5.5\times10^{-2}$ & $5.4\times10^{-1}$ & $5.4\times10^{0}$ & $5.8\times10^{-3}$ & $5.8\times10^{-2}$ & $5.7\times10^{-1}$ & $1.6\times10^{-8}$ & $1.5\times10^{-7}$ & $1.5\times10^{-6}$ & $3.0\times10^{-5}$ & $3.0\times10^{-4}$ & $2.9\times10^{-3}$\\ \hline
 %
 %
 {LT-ET-CE} & $1.5\times10^{-4}$ & $1.5\times10^{-3}$ & $1.5\times10^{-2}$ & $1.3\times10^{-4}$ & $1.3\times10^{-3}$ & $1.3\times10^{-2}$ & $4.7\times10^{-9}$ & $4.7\times10^{-8}$ & $4.6\times10^{-7}$ & $1.5\times10^{-5}$ & $1.5\times10^{-4}$ & $1.5\times10^{-3}$ \\ \hline
 %
 {AMIGO-ET-CE} & $4.9\times10^{-6}$ & $4.9\times10^{-5}$ & $4.8\times10^{-4}$ & $1.1\times10^{-4}$ & $1.1\times10^{-3}$ & $1.1\times10^{-2}$ & $2.4\times10^{-8}$ & $2.4\times10^{-7}$ & $2.4\times10^{-6}$ & $8.0\times10^{-6}$ & $7.9\times10^{-5}$ & $7.8\times10^{-4}$ \\ \hline 
%
 {LT-AMIGO-ET-CE} & $4.7\times10^{-6}$ & $4.6\times10^{-5}$ & $4.6\times10^{-4}$ & $1.1\times10^{-4}$ & $1.1\times10^{-3}$ & $1.1\times10^{-2}$ & $3.0\times10^{-9}$ & $2.9\times10^{-8}$ & $2.9\times10^{-7}$ & $6.2\times10^{-6}$ & $6.1\times10^{-5}$ & $6.1\times10^{-4}$\\ \hline
\end{tabular}
}
\begin{flushleft}
\footnotesize{Note: first column denotes the GW detector name or different combinations of them. Second, fourth and third columns show the low and high values of the $68\%$ confidence interval and the median value of the distribution of $\Delta\Omega_{90\%}$ among $100$ realizations of multiband BBHs. The fifth to seventh columns, the eighth to tenth columns and the last three columns show the corresponding results for the $\sigma_{d_{\rm L}}/d_{\rm L}$ distribution, the $\sigma_{\mc}/\mc$ distribution, and the $\sigma_{\eta}$ distribution, respectively.}
\end{flushleft}
\end{table*}

The bottom left panel of Figure~\ref{fig:f9} shows the localization errors and relative errors of luminosity distances of the multiband BBHs from the $50$th realization detected by AMIGO, ET-CE, and AMIGO-ET-CE, respectively. Similar to the results from the low- and high-frequency analysis, ET-CE has advantages in localization and luminosity distance measurements because of the high S/Ns for those sources in the high-frequency band. As seen from the bottom left panel of Figure~\ref{fig:f11}, AMIGO-ET-CE may localize $98\%$ of those multiband BBHs from all the $100$ realizations in the sky areas of 
$7$$\times$$10^{-7}$-$2\times10^{-3}$\,deg$^2$ and measure their luminosity distances with relative errors in the range of 
$8$$\times$$10^{-5}$ to $0.03$. The distributions of $\Delta \Omega_{90\%}$ and $\sigma_{d_{\rm L}}/d_{\rm L}$ by AMIGO-ET-CE have median values of $4.9$$\times$$10^{-5}$\,deg$^2$ and $1.1$$\times$$10^{-3}$, and their $68\%$ confidence intervals are $4.9$$\times$$10^{-6}$-$4.8\times10^{-4}$\,deg$^2$ and $1.1$$\times$$10^{-4}$-$0.011$, respectively. Comparing with the results obtained from the observations of AMIGO/ET-CE only, the improvements of the localization and luminosity distance measurement precision by AMIGO-ET-CE are 
$\sim 2$$\times$$10^{5}$/$120$ and $\sim 68$/$1.6$, respectively.

The bottom left panel of Figure~\ref{fig:f10} shows the estimation uncertainties about $\mathcal{M}_{\rm c}$ and $\eta$ for BBHs in the $50$th realization detected by AMIGO, ET-CE, and AMIGO-ET-CE, respectively. AMIGO dominates the estimation precision about $\mathcal{M}_{\rm c}$ and $\eta$ when considering the middle- and high-frequency multiband observations since the sources experience more cycles in middle frequency band. As shown in the bottom left panel of Figure~\ref{fig:f12}, AMIGO-ET-CE may measure $\mc$ for $98\%$ of those multiband BBHs among all $100$ realizations with relative errors in the range of $2$$\times$$10^{-8}$-
$3$$\times$$10^{-6}$,
and $\eta$ with errors in the range of $10^{-5}$-
$3$$\times$$10^{-4}$.
The distributions of $\sigma_{\mc}/\mc$ and $\sigma_{\eta}$ for multiband BBHs detected by AMIGO-ET-CE have median values of 
$2.4$$\times$$10^{-7}$ and $7.9$$\times$$10^{-5}$,
and their $68\%$ confidence intervals are 
$2.4$$\times$$10^{-8}$-$2.4$$\times$$10^{-6}$ and $8.0$$\times$$10^{-6}$-$7.9$$\times$$10^{-4}$, respectively. Compared with the results obtained from the observations by AMIGO or ET-CE only, the joint observations by AMIGO-ET-CE may improve the measurement precision of $\mathcal{M}_{\rm c}$/$\eta$ by a factor $\sim 2$ or $\sim 6700$/$\sim 6$ or $\sim 25$, respectively.

The bottom right panel of Figure~\ref{fig:f9} shows the estimation precision of sky localization and luminosity distances by the multiband observations combining low-, middle-, and high-frequency bands for those BBHs in the $50$th realization. LT-AMIGO-ET-CE gives the estimations of sky localization and luminosity distance measurements for the $18$ multiband BBHs a little bit better than those results given by AMIGO-ET-CE. The bottom right panel of Figure~\ref{fig:f11} shows that the combination of all three frequency bands gives the localization for $98\%$ of the multiband BBHs from all $100$ realizations in the sky areas of 
$7$$\times$$10^{-7}$-$2\times10^{-3}$\,deg$^2$ and the relative errors of luminosity distance in the range of 
$8$$\times$$10^{-5}$ to $0.03$.
The distribution of localization for those BBHs has the median value as 
$4.6$$\times$$10^{-5}$ and $68\%$ confidence interval of $4.7$$\times$$10^{-6}$-$4.6\times10^{-4}$.
Similarly, the median value and $68\%$ confidence interval of the $\sigma_{d_{\rm L}}/d_{\rm L}$ distribution are $1.1$$\times$$10^{-3}$ and $1.1$$\times$$10^{-4}$-$0.011$.
Compared with the results from AMIGO-ET-CE or LT-ET-CE, the localization can be improved by a factor $\sim 1.1$ or $\sim 32$, and the luminosity distance error can be improved by a factor $\sim 1.0$ or $\sim 1.2$ by LT-AMIGO-ET-CE. The combination of middle- and high-frequency information dominates the estimations of sky localization and luminosity distance errors since AMIGO measure the luminosity distance more accurate than LT does. 

The bottom right panel of Figure~\ref{fig:f10} shows the estimation precision of $\mc$ and $\eta$ for the multiband BBHs in the $50$th realization by the joint observations of all frequency bands. For the $18$ multiband BBHs, LT-AMIGO-ET-CE gives the relative errors of $\mc$ ranging from 
$5$$\times$$10^{-9}$ to $4$$\times$$10^{-8}$, and the errors of $\eta$ ranging from $10^{-5}$ to $2$$\times$$10^{-4}$. The bottom right panel of Figure~\ref{fig:f12} shows that LT-AMIGO-ET-CE may measure $\mc$ for $98\%$ of the multiband BBHs from all the $100$ realizations with relative errors in the range of $4$$\times$$10^{-9}$-$3$$\times$$10^{-7}$ and $\eta$ with errors in the range of $10^{-5}$-
$2$$\times$$10^{-4}$, respectively. The median value and $68\%$ confidence interval of the $\sigma_{\mc}/\mc$ (or $\sigma_{\eta}$) distribution obtained by the observations combining low-, middle-, and high-frequency bands are $2.9$$\times$$10^{-8}$ (or $6.1$$\times$$10^{-5}$) and $3.0$$\times$$10^{-9}$-$2.9$$\times$$10^{-7}$ (or $6.2$$\times$$10^{-6}$-$6.1$$\times$$10^{-4}$), respectively. These correspond to an improvement for the measurement of $\mc$ a factor $\sim 1.6$ or $\sim 8$ regarding to LT-ET-CE or AMIGO-ET-CE, and that for $\eta$ a factor $\sim 2$ or $\sim 1.3$ regarding to LT-ET-CE or AMIGO-ET-CE.

\section{Conclusions and Discussions}
\label{sec:dis_con}

In this paper, we investigate the prospects of GW multiband detection of those BBHs evolving from the low-frequency LISA-Taiji band through the middle-frequency AMIGO band to the high-frequency ET/CE band. We consider various formation models for BBHs and the constraint on the local merger rate density and its uncertainty given by the latest LIGO-VIRGO observations \citep{GWTC3population}. We estimate the number of detectable BBHs by LISA, Taiji, LT, bAMIGO/AMIGO/eAMIGO, and the multiband BBHs detected by the joint observations of low-, middle- and high-frequency GW detectors, which we define those sources with S/N both exceed $5$ at low- and middle-frequency band, via Monte Carlo realizations and simple S/N estimations. We also further estimate the precision of the localization and the physical parameter that can be obtained from the GW observations for these detectable BBHs.
 
We find that AMIGO in the middle-frequency band may detect $21-91$ BBHs with S/N $\geq 8$,  assuming that they operate in a continuous period of $4$ yr. The uncertainties in the estimated numbers are mainly due to the uncertainty in the constraint on the local merger rate density, while the choice of different BBH formation models only leads to a small change of the number if all these models are calibrated by the same local merger rate density. With lower/higher sensitivity, bAMIGO/eAMIGO may detect less/more ($1-13$/$121-454$) BBHs. AMIGO may be able to localize $98\%$ of the detectable BBHs among all $100$ realizations in sky areas of $\sim 0.02$-$5000$\,deg$^2$ and measure their luminosity distances, redshifted chirp masses, symmetric mass ratios with relative errors in the ranges of $0.01$-$0.5$, $6$$\times$$10^{-8}$-$9\times10^{-5}$, and $10^{-4}$-$3$$\times$$10^{-3}$, respectively.
Compared with AMIGO and bAMIGO, eAMIGO can achieve better estimation precision for those BBHs, which can also be detected by AMIGO and bAMIGO, because eAMIGO is more sensitive. However, the localization and the precision of other parameter estimations for the faraway sources, which can only be detected by eAMIGO, may be much poorer than those for AMIGO and bAMIGO detectable sources. Even with the observations of bAMIGO only, the localization of its detectable BBHs can be quite precise because they are all at substantially smaller distances. 

The joint observations of the low-frequency-middle-frequency GW network composed by LT and AMIGO (LT-AMIGO) may detect $5$-$33$ BBHs. Considering all the $100$ realizations with $\Rmrgbh = 19.1\gpcyr$, LT-AMIGO may provide the localization estimates for these multiband BBHs with its distribution median value as $0.54$\,deg$^2$ and $68\%$ confidence interval as $0.055$-$5.4$\,deg$^2$, which is about a factor of $1.5$ or $20$ better than those obtained by using the observations of LT or AMIGO only. 
It can also help to measure $\sigma_{d_{\rm L}}/d_{\rm L}$, $\sigma_{\mc}/\mc$, and $\sigma_{\eta}$ with their distribution median values as $0.058$, $1.5$$\times$$10^{-7}$, and $3.0$$\times$$10^{-4}$, respectively. The corresponding $68\%$ confidence intervals for those distributions are $5.8$$\times$$10^{-3}$-$0.57$, $1.6$$\times$$10^{-8}$-$1.5$$\times$$10^{-6}$, and $3.0$$\times$$10^{-5}$-$2.9$$\times$$10^{-3}$, which are about a factor of $2$ (or $1.3$), $21$ (or $3$), and $18$ (or $1.6$) better than those measured by only using the observations by LT (or AMIGO).

The joint observations by LT-ET-CE may localize the detectable BBHs with its distribution median value as $1.5\times10^{-3}$\,deg$^2$ and the corresponding $68\%$ confidence interval as $1.5$$\times$$10^{-4}$-$0.015$\,deg$^2$. It helps constraining $\sigma_{d_{\rm L}}/d_{\rm L}$, $\sigma_{\mc}/\mc$, and $\sigma_{\eta}$ with their distribution median values as $1.3$$\times$$10^{-3}$, $4.7$$\times$$10^{-8}$, and $1.5$$\times$$10^{-4}$, respectively. The corresponding $68\%$ confidence intervals for those distributions are $1.3$$\times$$10^{-4}$-$0.013$, $4.7$$\times$$10^{-9}$-$4.6$$\times$$10^{-7}$, and $1.5$$\times$$10^{-5}$-$1.5$$\times$$10^{-3}$, respectively. Comparing with the observations by LT or ET-CE only, the improvements by joint LT-ET-CE observations are a factor of $\sim 550$ or $\sim 4$, $\sim 84$ or $\sim 1.3$, $\sim 72$ or $\sim 3\times10^4$, and $\sim 37$ or $\sim 13$ for the localization, and measurement precision of the luminosity distance, chirp mass, and symmetric mass ratio, respectively.

The joint observations by AMIGO-ET-CE may localize the detectable multiband BBHs with the distribution median value as $4.9$$\times$$10^{-5}$\,deg$^2$ and the corresponding $68\%$ confidence interval as $4.9$$\times$$10^{-6}$-$4.8\times10^{-4}$\,deg$^2$. AMIGO-ET-CE helps determining $\sigma_{d_{\rm L}}/d_{\rm L}$, $\sigma_{\mc}/\mc$, and $\sigma_{\eta}$ with their distribution median values as $1.1$$\times$$10^{-3}$, $2.4$$\times$$10^{-7}$, and $7.9$$\times$$10^{-5}$, respectively. The corresponding $68\%$ confidence intervals for those distributions are $1.1$$\times$$ 10^{-4}$-$0.011$, $2.4$$\times$$10^{-8}$-$2.4$$\times$$10^{-6}$, and $8.0$$\times$$10^{-6}$-$7.9$$\times$$10^{-4}$, respectively. Comparing with those given by the observations of AMIGO or ET-CE only, the improvements by AMIGO-ET-CE joint observations are a factor of $\sim 2$$\times$$10^{5}$ or $\sim 120$, $\sim 68$ or $\sim 1.6$, $\sim 2$ or $\sim 6700$, and $\sim 6$ or $\sim 25$ for the localization, and precision of the luminosity distance, chirp mass, and symmetric mass ratio measurements, respectively.

The combination of all three band observations by LT-AMIGO-ET-CE leads to further improvements on the localization and parameter estimation of those multiband detectable BBHs. The median value and $68\%$ confidence interval of the distribution of sky localization by LT-AMIGO-ET-CE are $ {4.6}$$\times$$ {10^{-5}}$\,deg$^2$ and $ {4.7}$$\times$$ {10^{-6}}$-$ {4.6\times10^{-4}}$\,deg$^2$, which is about a factor of $\sim 1.1$ or $\sim  {32}$ better than those by AMIGO-ET-CE or LT-ET-CE. The median values of the distributions for $\sigma_{d_{\rm L}}/d_{\rm L}$, $\sigma_{\mc}/\mc$, and $\sigma_{\eta}$ are $ {1.1}$$\times$$ {10^{-3}}$, $ {2.9}$$\times$$ {10^{-8}}$, and $ {6.1}$$\times$$ {10^{-5}}$, respectively. The corresponding $68\%$ confidence intervals are $ {1.1}$$\times$$ {10^{-4}}$-$ {0.011}$, $ {3.0}$$\times$$ {10^{-9}}$-$ {2.9}$$\times$$ {10^{-7}}$, and $ {6.2}$$\times$$ {10^{-6}}$-$ {6.1}$$\times$$ {10^{-4}}$, respectively. Compared with those by the observations of AMIGO-ET-CE (or LT-ET-CE), LT-AMIGO-ET-CE gives the measurement accuracy for luminosity distance, redshifted chirp mass, and symmetric mass ratio an improvement by a factor $\sim  {1.0}$ (or $\sim  {1.2}$), $\sim  8$ (or $\sim  {1.6}$), and $\sim  {1.3}$ (or $\sim 2$), respectively. A significant fraction of the LT-AMIGO-ET-CE multiband BBHs can be localized within sky localization of $10^{-3}$\,deg$^2$ and have distance measurements with relative errors of $\sigma_{d_{\rm L}}/d_{\rm L} \lesssim 0.001$, of which the host galaxies may be identified directly by the GW observations.

The GW observations in different frequency bands may exhibit different advantages in the parameter estimations for multiband BBHs. For example, LT may localize a source better than AMIGO does since it has longer baseline and those sources stay more time in the lower frequency band. High-frequency detectors, like ET-CE, achieve the best localization ability because the S/N of the source detected by ET-CE is much higher than that by LT or AMIGO. As for the measurement precision of $\mc$ and $\eta$, AMIGO gives smaller errors than  LT does because of the faster frequency evolution in the middle-frequency band, and low-/middle-frequency detectors are better in estimating these two parameters than ET-CE since the sources stay in low- and middle-frequency band experiencing more cycles. Significant improvement in the estimations of BBH physical parameters and localization suggests that the multiband GW observations are  helpful for testing the theory of general relativity, probing the nature of gravity, and inferring the cosmological parameters.

We note that several approximations made in this paper may affect the estimates quantitatively though do not change the general results. For rough estimation of the multiband GW detection, only simplified cases are considered. In reality, there are many complexities that one should take into account to make a more robust investigation. In this paper, we simply set the S/N threshold as the typical value for BBH detection $\varrho=8$ in each single frequency band, while set $\varrho=5$ in low-, middle-, and high-frequency band at the same time for detectable multiband BBHs, which are however arguable for a few reasons. First, the FIM method is adopted rather than a full Bayesian approach \citep[e.g.][]{2014ApJ...789L...5K, 2014PhRvD..89d2004G, 2015PhRvD..91d2003V}, to estimate the improvement of the accuracy of parameter determination by multiband GW observations. However, this method is only valid under the high-S/N approximation when the contributions of non-linear terms in the signals can be neglected \citep[][]{2008PhRvD..77d2001V}. Roughly speaking, the FIM method may introduce one order of magnitude overestimation on the localization precision of a GW source with a relatively low S/N ($\varrho < 8$) \citep[e.g.,][]{2013PhRvD..88h4013R,2022arXiv220702771I}. Second, due to the large number of templates required for the data processing in the low-frequency GW detection, a higher S/N may be also needed, for example, $\varrho=15$ as suggested by \citet{Moore2019}. If adopting this value, the numbers estimated in this work may decrease by a factor of $\sim 5$ to one order of magnitude. 

We assume that LISA, Taiji, and AMIGO start observation at the same time and all have $4$ yr' continuous observation period, which may be not the same as the real situation in the future. If AMIGO starts its observation after the end of observation of LISA/Taiji, some of the BBHs merging within the observation period of LISA/Taiji analysed above will not be observed by AMIGO. However, some of the BBHs with coalescence time-scale larger than $4$ yr detected in the low-frequency band may evolve to the middle-frequency band eventually and be observed by AMIGO. Nevertheless, our investigations on the detectable BBHs by different GW detectors at different frequency band can be taken as an illustration to multiband GW observations.

We note here that the low-frequency and middle-frequency GW observations may be helpful for measuring the eccentricities of BBHs \citep{Nishizawa2016,Nishizawa2017,Manuel2020} and the high-frequency GW observations can provide census on the effective spin distribution of those BBHs \citep{Farr2017,Roulet2021}. Therefore, the multiband GW observations by combining the GW detectors in different frequency band can help to distinguish different BBH formation models or constrain the contribution fractions by different models. For demonstration purpose of this paper, however, BBHs possessing spin and non-circular orbits are not taken into consideration. We defer the investigation on the non-circular spinning BBHs for future study.

\section*{Acknowledgements}
We thank Yunfeng Chen and Xiao Guo for helpful discussions. This work is partly supported by the National Key Program for Science and Technology Research and Development (Grant No. 2020YFC2201400), the National Natural Science Foundation of China (Grant No. 11690024, 12273050, 11873056, 11991052), the Strategic Priority Program of the Chinese Academy of Sciences (Grant No. XDB 23040100).

\section*{Data Availability}

The data used in this paper will be shared on reasonable request to the corresponding author.



\bibliographystyle{mnras}
\bibliography{example} 








\bsp	
\label{lastpage}
\end{document}